\documentclass[sigconf]{acmart}
\usepackage{algorithm}
\usepackage{algorithmic}
\usepackage{amsmath}
\usepackage{amsfonts}
\usepackage{comment}
\usepackage{multirow}
\usepackage{tabularx}
\usepackage{colortbl}
\usepackage{pdflscape}
\usepackage{array}
\usepackage{ragged2e}
\usepackage{makecell}
\usepackage{bigstrut}
\usepackage{soul}
\usepackage[most]{tcolorbox}
\usepackage{stfloats}
\definecolor{skyblue}{rgb}{0.53, 0.81, 0.98}
\definecolor{mycustomcolor}{HTML}{F5C4F5}
\newtcbox{\redhl}{on line,
  arc=3pt,outer arc=3pt,
  colback=red!30!white,colframe=red!50!black,
  boxsep=0pt,left=2pt,right=2pt,top=2pt,bottom=2pt,
  boxrule=0pt}

\newtcbox{\pinkhl}{on line,
  arc=3pt, outer arc=3pt,
  colback=mycustomcolor, colframe=pink!50!black,
  boxsep=0pt, left=2pt, right=2pt, top=2pt, bottom=2pt,
  boxrule=0pt  
}

\newtcbox{\greenhl}{on line,
  arc=3pt,outer arc=3pt,
  colback=green!20!white,colframe=green!50!black,
  boxsep=0pt,left=2pt,right=2pt,top=2pt,bottom=2pt,
  boxrule=0pt}

\newtcbox{\bluehl}{on line,
  arc=3pt,outer arc=3pt,
  colback=skyblue!90!white,colframe=skyblue!95!black,
  boxsep=0pt,left=2pt,right=2pt,top=2pt,bottom=2pt,
  boxrule=0pt}

\newtcbox{\yellowhl}{on line,
  arc=3pt,outer arc=3pt,
  colback=yellow!30!white,colframe=yellow!50!black,
  boxsep=0pt,left=2pt,right=2pt,top=2pt,bottom=2pt,
  boxrule=0pt}

\newtcbox{\grayhl}{on line,
  arc=3pt,outer arc=3pt,
  colback=gray!10!white,colframe=gray!50!black,
  boxsep=0pt,left=2pt,right=2pt,top=2pt,bottom=2pt,
  boxrule=0pt}

\AtBeginDocument{%
  \providecommand\BibTeX{{%
    \normalfont B\kern-0.5em{\scshape i\kern-0.25em b}\kern-0.8em\TeX}}}

\newcolumntype{L}[1]{>{\raggedright\arraybackslash}m{#1}}
\definecolor{customcolor1}{RGB}{189,189,189}
\definecolor{customcolor2}{RGB}{198,218,242}

\definecolor{customcolor3}{RGB}{234,246,246}
\definecolor{customcolor4}{RGB}{245,248,253}

\definecolor{myhighlight}{rgb}{0.8, 1, 0.8}
\sethlcolor{myhighlight}
\AtBeginDocument{%
  \providecommand\BibTeX{{%
    \normalfont B\kern-0.5em{\scshape i\kern-0.25em b}\kern-0.8em\TeX}}}

\copyrightyear{2024} 
\acmYear{2024} 
\setcopyright{rightsretained} 
\acmConference[CHI '24]{Proceedings of the CHI Conference on Human Factors in Computing Systems}{May 11--16, 2024}{Honolulu, HI, USA}
\acmBooktitle{Proceedings of the CHI Conference on Human Factors in Computing Systems (CHI '24), May 11--16, 2024, Honolulu, HI, USA}
\acmDOI{10.1145/3613904.3642117}
\acmISBN{979-8-4007-0330-0/24/05}

\acmConference[CHI ’24]{Make sure to enter the correct
  conference title from your rights confirmation emai}{May 11--16,
  2024}{Hawaii, USA}
\acmBooktitle{CHI ’24, May 11--16, 2024  Hawaii, USA} 
\acmPrice{15.00}
\acmISBN{978-1-4503-XXXX-X/18/06}
\usepackage{soul,color}

\begin{document}

\title[Predicting Pandemic Health Decisions and Outcomes Through Social Media and LLMs]{Leveraging Prompt-Based Large Language Models: Predicting Pandemic Health Decisions and Outcomes Through Social Media Language}
\author{Xiaohan Ding}
\orcid{0009-0003-2679-3344}
\affiliation{%
  \institution{Department of Computer Science, Virginia Tech}
  \city{Blacksburg}
  \country{USA}}
\email{xiaohan@vt.edu}

\author{Buse Carik}
\affiliation{%
  \institution{Department of Computer Science, Virginia Tech}
    \city{Blacksburg}
  \country{USA}}
\email{buse@vt.edu}

\author{Uma Sushmitha Gunturi}
\orcid{0000-0002-2045-0792}
\affiliation{%
  \institution{Department of Computer Science, Virginia Tech}
    \city{Blacksburg}
  \country{USA}}
\email{umasushmitha@vt.edu}

\author{Valerie Reyna}
\affiliation{%
  \institution{Human Neuroscience Institute, Cornell University}
  \city{New York}
  \country{USA}}
\email{vr53@cornell.edu}

\author{Eugenia H. Rho}
\affiliation{%
  \institution{Department of Computer Science, Virginia Tech}
    \city{Blacksburg}
  \country{USA}}
\email{eugenia@vt.edu}

\renewcommand{\shortauthors}{Ding et al.}

\begin{abstract}
We introduce a multi-step reasoning framework using prompt-based LLMs to examine the relationship between social media language patterns and trends in national health outcomes. Grounded in fuzzy-trace theory, which emphasizes the importance of “gists” of causal coherence in effective health communication, we introduce Role-Based Incremental Coaching (RBIC), a prompt-based LLM framework, to identify gists at-scale. Using RBIC, we systematically extract gists from subreddit discussions opposing COVID-19 health measures (Study 1). We then track how these gists evolve across key events (Study 2) and assess their influence on online engagement (Study 3). Finally, we investigate how the volume of gists is associated with national health trends like vaccine uptake and hospitalizations (Study 4). Our work is the first to empirically link social media linguistic patterns to real-world public health trends, highlighting the potential of prompt-based LLMs in identifying critical online discussion patterns that can form the basis of public health communication strategies.

\end{abstract}


\begin{teaserfigure}
  \includegraphics[width=0.93\textwidth]{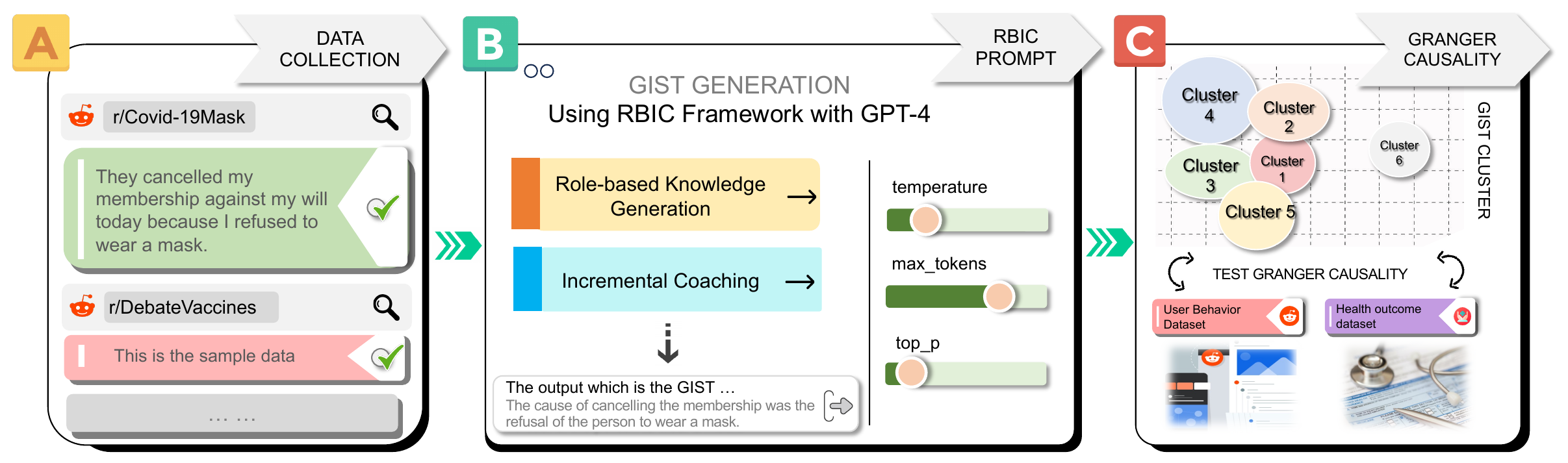}
  \caption{Our research empirically links social media conversations to national health outcomes related to COVID-19. We introduce Role-Based Incremental Coaching (RBIC), a Large Language Model (LLM) prompting framework. (A) Collecting Reddit datasets focused on communities known for opposing COVID-19 health practices. (B) Guided by Fuzzy-Trace Theory, we introduce a novel LLM framework called Role-Based Incremental Coaching (RBIC) to extract cause-effect pairs and formulate coherent gists capturing causal relations in texts. (C) Granger-causality tests and data analytics reveal the impact of these gists on community engagement and national health outcomes such as vaccination uptake and hospitalization rates.}
  \label{fig:teaser}
\end{teaserfigure}
\maketitle

\section{Introduction}
During the COVID-19 pandemic, social media was at the center of proliferating mass antipathy and distrust towards government health policies and recommendations \cite{Demuyakor2021Unmasking, social2023}. Millions took to social media to oppose federal and state health practices, criticize medical professionals, or organize anti-vaccine and mask-wearing rallies \cite{martin_any_2022}. The viral growth of such online conversations fueled animosity and extremist views that encouraged people to resist public health guidelines \cite{Demuyakor2021Unmasking, Al-Ramahi2020Public}. Disregarding public health practices, such as wearing masks, maintaining social distance, and getting vaccinated resulted in significant societal costs. Between November and December of 2021 alone, over 692,000 preventable hospitalizations were reported among unvaccinated patients, leading to a staggering \$13.8 billion \cite{farrenkopf_cost_2022}. Soaring COVID-19 infection cases put a massive burden on healthcare systems, depleting medical resources and contributing to severe employee burn-outs and shortages of healthcare workers \cite{kretchy_medication_2021}. Meanwhile, COVID-19 conspiracies and hyper-partisan news on social media led to nationwide protests, obstruction of medical facilities \cite{baker_trust_2020}, and even fatal assaults of employees requesting customers to wear masks \cite{bhattarai_retail_2020}. 

According to fuzzy-trace theory (FTT), texts that clearly establish cause-and-effect relationships facilitate humans extraction of \textit{gist} mental representations, helping people understand and remember information better than texts without any causal coherence \cite{reyna_new_2012,reyna_how_2016}. This aligns with previous studies in decision sciences, which have shown that causal coherence of \textit{gists} in texts plays a crucial role in how individuals perceive risks and make health-related decisions \cite{corbin_how_2015, reyna_supporting_2022}. Throughout the pandemic, social media conversations refuting COVID-19 public health practices based on mis-/disinformation and identity politics continued to obscure people’s knowledge of safe health practices, making well-informed health decisions extremely difficult \cite{weinzierl_identifying_2022}. Using evidence-based theories like FTT allows us to create psychologically descriptive models that transform language into analyzable units shown to predict human behavior \cite{Valerie-pnas, corbin_how_2015}.

In this paper, we leverage the capabilities of prompt-based Large Language Models (LLMs) to delve into the nuanced language patterns in social media discussions opposing COVID-19 public health practices through a theory-driven approach using fuzzy-trace theory. By leveraging prompt-based LLMs, we dissect the language around resistance to COVID-19 health practices through the lens of FTT and its central concept of \textit{gist}. Specifically, we examine how causal language patterns or \textit{gists} manifest across social media communities that denounce pandemic health practices, contribute to trends in people's health decisions, and by extension, impact national health outcomes. We divide our work into four main studies to address the following research questions:

\begin{itemize}
    \item RQ1. How can we efficiently predict gists across social media discourse at-scale? (Study 1)
    \item RQ2. What kind of gists characterize how and why people oppose COVID-19 public health practices? How do these gists evolve over time across key events? (Study 2)
    \item RQ3. Do gist patterns significantly predict patterns in online engagement across users in banned subreddits that oppose COVID-19 health practices? (Study 3)
    \item RQ4. Do gist patterns significantly predict trends in national health outcomes? (Study 4)
\end{itemize}

We answer RQ1 by leveraging LLMs and their prompt-based capabilities to identify gists in social media conversations at-scale (Study 1). We do so by developing a novel prompting framework that detects and extracts cause-effect pairs in sentences from a corpus of online discussions collected from banned Reddit communities known for opposing public COVID-19 health practices. Study 1 allows us to identify the causal language (cause-effect pairs that form \textit{gists}) that underlie how people argue against COVID-19 health practices on social media. We answer RQ2 by clustering sentence embeddings of gists (sentences with causal relations identified from Study 1) to identify the most salient gist clusters, and demonstrate how they evolve across key events (Study 2). Finally, we answer RQs 3 and 4 by using Granger-causality to test whether causal discourse (gists) on social media can significantly predict online engagement patterns (Study 3), and trends in national health decisions and outcomes in the U.S. (Study 4). 

\textbf{Contributions}: This work's intellectual merits are methodological and theoretical. The computational techniques introduced in this work enable efficient and scaled prediction of gists on social media, and thus can be used to better identify and understand underlying mental representations that motivate health decisions and attitudes towards public health practices (Study 1). The clustering and evolution of gists in Study 2 identify the most salient themes associated with how people causally argue against pandemic health practices online. Patterns in gist volumes across cluster topics fluctuate closely with topically-related high-profile events, including federal health announcements, congressional policies, and remarks by a country’s leader. Study 3 empirically confirms how gist volumes significantly drive subreddit engagement patterns (upvotes and comments), providing implications for how causal language may play a role in monitoring conversations in content-moderation practices of controversial online health communities. Finally, gist patterns within subreddits that support anti-pandemic health practices were significantly interrelated with nationwide trends in important health decisions and outcomes (Study 4). To the best of our knowledge, our research is the first to empirically establish Granger causality between linguistic patterns in social media discussions about COVID-19 health measures and real-world trends in public health outcomes. Our work entails the following contributions:  
\begin{itemize}
       \item The task of accurately predicting causal language patterns and generating coherent gists (causal statements) is a complex challenge \cite{Gist-and-Verbatim, reyna2023numeracy}. We overcome this by introducing a multi-step prompting framework: Role-Based Incremental Coaching (RBIC). RBIC is a prompting mechanism that allows efficient prediction of gists across social media conversations at-scale. RBIC integrates role-based cognition with effective learning in sub-tasks to enhance the model's overall understanding of a given task prior to generating a final output. We overcome prior challenges in detecting subtle and complex expressions of semantic causality in noisy text by leveraging RBIC. By doing so, this work advances state-of-the-art approaches in detecting gists at-scale, yielding a novel, psychologically relevant, and efficient technique for identifying and examining bottom-line meanings in massive amounts of textual data. 

       \item We demonstrate the novel application of prompt-based LLMs in advancing computational social science (CSS) methods in Human-Computer Interaction (HCI) research. Generic Natural Language Processing (NLP) models and LLMs typically lack multi-step reasoning capabilities \cite{AIChains}. This limitation makes it difficult to apply such models in performing nuanced and complex text analyses in CSS research \cite{ziems_can_2023}.  By applying RBIC, we overcome this limitation and demonstrate the versatility and effectiveness of prompt-based LLMs in identifying and synthesizing nuanced linguistic patterns. In so doing, we contribute to broadening the potential application of prompt-based LLMs for theory-driven textual analysis in CSS research in the HCI domain. 

    \item Our research enhances the analytical depth and scope of insights into the causal discourse surrounding people’s opposition to public health practices on social media. We identify the most salient gist clusters that embody the core topics at the center of how and why people oppose public health practices throughout COVID-19, from May 2020 to October 2021. We use sentence embeddings and clustering to provide a characterization of how the volume of gists across each topic fluctuates in relation to key events associated with the core topics embodied by the gist clusters. By doing so, we capture how causal online discourse surrounding anti-COVID-19 health practices evolves over time across real-world events. Such insights can, in turn, inform timely public health communication strategies and interventions that account for ongoing current events \cite{Valerie-pnas}.

    \item Finally, we address the question of whether and how social media language patterns in the form of gists influence nationwide trends in vaccinations, COVID-19 cases, and hospitalization in the U.S., providing new evidence around how important health decisions and national health outcomes are impacted by causal linguistic signatures across social media health discussions—an important link that has not been empirically established at-scale in prior research.
\end{itemize}
\section{Related Work}
\subsection{Understanding the Impact of Social Media Language Patterns on Health Decisions and Outcomes }

The COVID-19 pandemic has ignited an unprecedented increase in social media discourse on health decisions and practices \cite{zhang_distress_2021, weinzierl_identifying_2022}, spurring a wave of computational social science research \cite{trajkova_exploring_2020, Mental-Health} aimed at understanding this phenomenon in the field of HCI \cite{covid-hci,covid-hci-ihc} and CSCW \cite{caldeira_crisis_2023}. Using text mining and computational linguistics, researchers have analyzed pandemic-related social media discourse through the lens of mental health \cite{10.1145/2740908.2743049},  political views \cite{Political_Ideology_Predicts, rheault_cochrane_2020}, attitudes towards vaccines \cite{Text_Mining_Anti, qorib_covid-19_2023}, misinformation \cite{su_motivations_2020, 10.1145/3491102.3517622,10.1145/3578503.3583617}, and perceptions of health policies and government institutions \cite{hussain_opportunities_2021}. Such studies have uncovered key insights on how language patterns reflect people’s beliefs \cite{Health_Beliefs_covid}, sentiments \cite{Sentiment_covid}, and emotional well-being \cite{Well-Being-covid, survey_health_behaviors} during Covid-19. For example, researchers have examined collective shifts in the public mood in response to the evolving pandemic news cycles by analyzing the daily sentiment of tweets \cite{Longitudinal-Analysis}. Similarly, others have analyzed social media posts containing a subset of depression-indicative n-grams to track the fluctuation in mental health of social media users over the course of the pandemic \cite{Mental-Health}. 

While such studies have made valuable contributions to understanding the role of language patterns in health-related discourse on social media \cite{chi23-Health-Discourse,10.1145/3458770}, there remains an opportunity to explore their impact on real-world health decisions and outcomes. To the best of our knowledge, there has been a lack of research that examines how social media discussion patterns surrounding health practices can predict patterns in health decisions and outcomes in the real world. Our research aims to fill this gap. Some emerging research, such as the study by Nyawa et al. (2022), has started to explore this link by applying computational linguistics to categorize individuals as either vaccine-accepting or vaccine-hesitant based on their online language patterns \cite{nyawa_covid-19_2022}. Yet, the majority of empirical studies examining the impact of social media discourse on real-world behavior thus far have leaned heavily on survey-based methods \cite{survey_vaccine_hesitancy,survey_health_behaviors}. These surveys often depend on self-reported metrics about social media use and health behaviors, thereby offering only a limited perspective on the complex relationship between social media discourse patterns and actual health decisions. This limitation underscores the existing challenges in understanding how health-related discussions on the internet translate into or shape real-world outcomes and decisions \cite{bavel_using_2020}.  Our research aims to address this challenge by investigating how language patterns in social media conversations can serve as predictive markers for understanding real-world trends in people’s health decisions and outcomes during the pandemic.

\subsection{Understanding Health Discourse Through the Lens of Fuzzy-Trace Theory and Its Core Concept of Gist}

Scholars have used fuzzy-trace theory (FTT) as a theoretical lens to explore risk perceptions and decisions underlying health practices and discussions in various contexts, including vaccines \cite{weinzierl_identifying_2022}, cancer \cite{wolfe_efficacy_2015}, HIV-AIDS \cite{wilhelms_gist_2015} and the prescription of antibiotics \cite{klein_categorical_2017}. These studies support FTT’s core tenet that gists are stronger and more effective forms of communication than verbatim representations in the sense that they are (a) better remembered and (b) more likely to influence decisions \cite{reyna_new_2012,reyna_how_2016}.
For example, a study comparing articles on vaccines posted on Facebook showed that those containing gists (e.g., bottom-line meaning) are shared 2.4 times more often on average than articles with verbatim details (e.g., statistics) \cite{broniatowski_effective_2016}. Having a story or images did not add unique variance to predictions once gist was accounted for. The study’s results show that communications about vaccines are more widespread when they express a clear gist explaining the bottom-line meaning of the statistics rather than just the data themselves. Likewise, scholars have also used FTT as a theoretical framework to examine people's behavior across diverse contexts, such as law, medicine, public health, systems engineering, and HCI \cite{marti_does_2020, reyna_supporting_2022,zottoli_developing_2023}.
For example, in HCI, researchers have used FTT to examine people's behavior in online social tagging \cite{Tagging} and to improve speech-to-text interface design through gist-based communication \cite{Gist-and-Verbatim}. Others have used FTT in designing a web-based intelligent tutoring system for communicating the genetic risk of breast cancer through gists \cite{wolfe_efficacy_2015}. 
Overall, FTT’s theoretical breadth and empirical support as a cognitive explanation of how people process and communicate information related to health decisions makes FTT a well-suited theoretical lens to examine resistance towards public health practices in our research. 

Further, gists that causally link some event, actor, or outcome tend to facilitate more effective uptake of information than those that are less causally coherent \cite{Valerie-pnas,liederholm2000effects}. In fact, causal coherence is one of the most important semantic aspects of gists that make gist-based communications effective \cite{hosseini_does_2019}. For example, in a study analyzing 9,845 vaccine-related tweets, researchers discovered that tweets containing explicitly causal gists (e.g., "vaccines cause autism") were far more likely to be retweeted and to go viral. This was in contrast to tweets that suggested a link between vaccines and autism but emphasized details and lacked a meaningful causal connection \cite{broniatowski_effective_2016}. Simply, information with stronger causal structure produces more meaningful gists in people, who then are more likely to remember, apply, and share that information \cite{reyna2023numeracy}. Fuzzy-trace theory draws on psycholinguistic research on mental representations of narratives that underlies both human memory models and computational models in which causal connections are a central feature of common gists \cite{Reyna1994DevelopmentOG, Trabasso1985CausalTA}.
Hence, we focus on causal gists, or gists that contain a cause-effect relation. From hereon, we refer to causal gists as gists.

\subsection{Challenges in Predicting Semantic Causality in Online Health Discourse }
Extracting cause-effect relations in text is one of the many open challenges in NLP research that has seen significant breakthroughs in recent years through the development of generative Large Language Models \cite{survey_yang_2022}. However, computational social science research has yet to take advantage of these advancements \cite{ziems_can_2023}, particularly in examining gists related to health practices. For example, scholars have used topic modeling, such as Latent Dirichlet Allocation (LDA) \cite{blei_latent_2003} to identify gists in vaccine hesitancy \cite{hosseini_does_2019}. While useful, these methods do not enable granular detection of gists at the sentence or phrase level. For instance, LDA only allows the detection of gists at the corpus level, where each identified topic across the entire dataset is treated as a proxy identification of one gist. Recent scholarship in medical informatics has examined health-related attitudes in social media by extracting causality through machine learning approaches with rule-based dependency parsing and named entity recognition \cite{mihaila_semi-supervised_2014,mirza-tonelli-2016-catena, Prasad2019ThePD, caselli-vossen-2017-event, doan_extracting_2019}. While such approaches are an improvement, they can only detect intra-sentential (within a single sentence) and not inter-sentential causality where cause and effect lie in different sentences (e.g., God made us to breathe naturally. I won’t be forced to wear masks.). More recently, transformer models such as InferBERT and CausalBERT, specifically designed for extracting causal relationships, have yielded more promising results \cite{wang_inferbert_2021, khetan_causal_2021}. However, the token limit of these models significantly reduces performance when dealing with longer texts \cite{ali_towards_2022}. Additionally, like humans, these models struggle to discern subtle forms of semantic causality in noisy or incoherent data. Our research aims to not only identify causality in text, but also generate coherent gists based on the identified cause-effect pairs. To achieve this, we address prior limitations by leveraging recent advancements in pretrained LLMs and their prompt-based approaches to develop a novel prompting framework to systematically predict gists \cite{wei_chain--thought_2023}.

\section{Study 1: Predicting Gists in Social Media Conversations at-Scale}

\begin{figure*}[!ht]
\centering
\includegraphics[width=2.1\columnwidth]{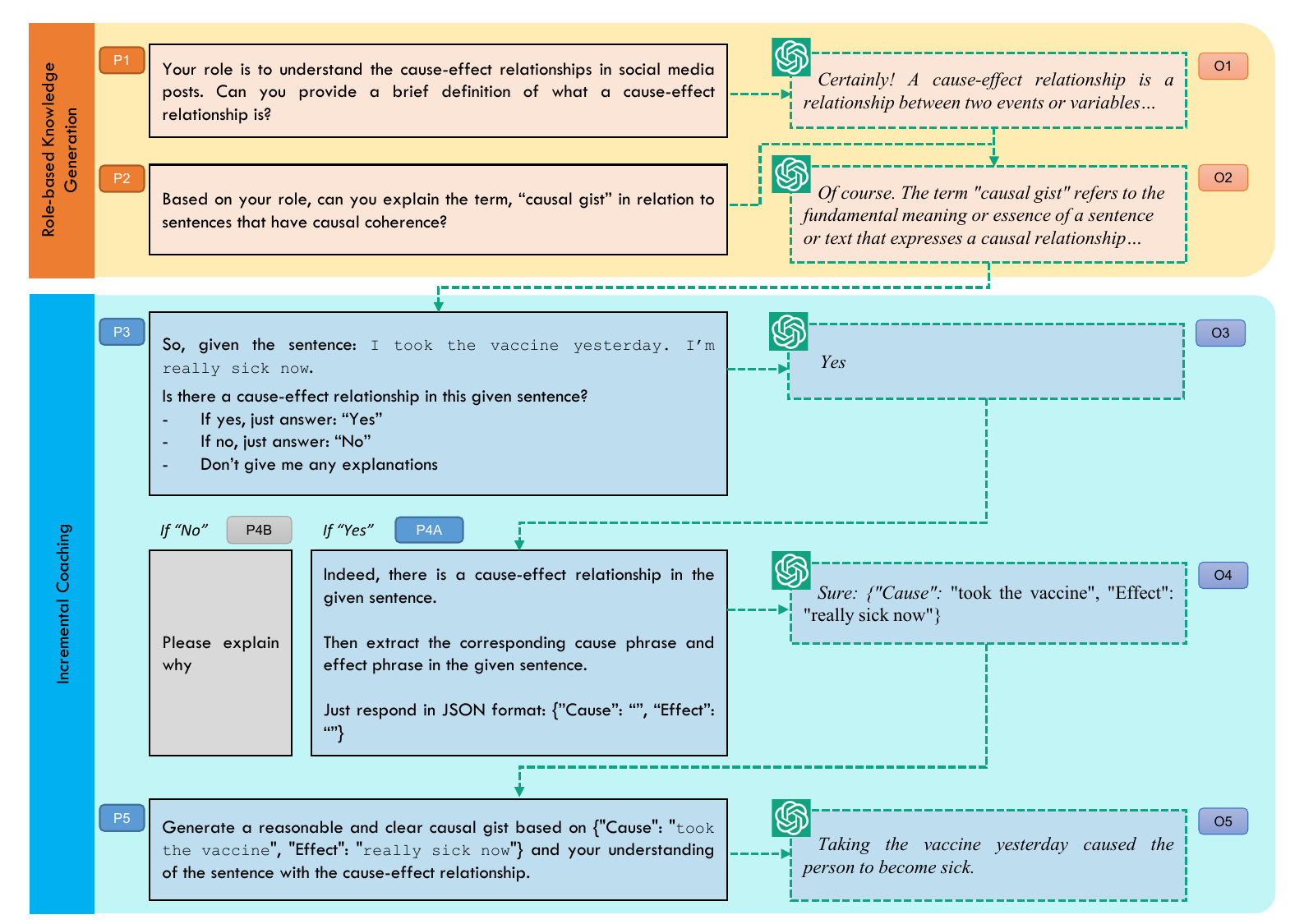}
\caption{Illustration of the Role-Based Incremental Coaching (RBIC) prompting framework: RBIC incorporates role-based cognition and sub-task training to improve the model's comprehension of a specific task before generating the final output.}
\label{rq1-workflow}
\end{figure*}

As a first step to analyzing how causal language patterns on social media impact health decisions and outcomes, we leverage the power of prompt-based LLMs in Study 1. Specifically, we develop and apply a multi-step prompting framework called \textit{Role-Based Incremental Coaching (RBIC)} to \textbf{efficiently predict gists across social media discourse at-scale}. Role-Based Incremental Coaching is a prompting framework (Fig. \ref{rq1-workflow}) built with few-shot demonstrations using GPT-4, which consists of two primary prompting techniques: \textit{Role-Based Knowledge Generation} and \textit{Incremental Coaching}. Combined together, RBIC allows the model to  (1) learn its role for a given task by generating role-specific knowledge as a task-performing agent and (2) perform a series of small sub-tasks to refine its understanding and quality of the final output by incrementally building upon the sub-task responses. RBIC allows us to systematically identify the presence of semantic causality in a given post, and generate causally coherent gists across large volumes of textual corpora at-scale. 
\subsection{Data}  

We collected all publicly available posts from 20 anti-COVID-19 subreddits that were banned for denouncing COVID-19 public health practices. These subreddits were chosen based on their community size, as well as the significant media attention they received from major news outlets \cite{CNN_Iyengar_2021}, and their virality among American social media users \cite{NYT_Isaac_2020}. We obtained all posts and corresponding metadata (comments, post id, timestamp, up/down-vote ratio, etc.) for each of these subreddits using Pushift API. This resulted in a total of 79,680 posts spanning from May 2020 to October 2021 from the following subreddits: \textit{conspiracy\_commons, CoronavirusCirclejerk, CoronavirusFOS, Coronavirus\_Rights, COVID19, covid19\_testimonials, covidvaccinateduncut, VaxKampf, DebateVaccines, FauciForPrison, ivermectin, lockdownskepticism, NoNewNormal, trueantivaccination, vaccinelonghaulers, VAERSreports, Wuhan\_Flu, CovidIsAFraud, COVID19Origin, churchofcovid.}

\subsection{Method: Role-Based Incremental Coaching (RBIC)}
\textbf{Role-Based Knowledge Generation}. Drawing inspiration from prior NLP research that leverages multi-step reasoning capabilities in LLMs \cite{liu_generated_2022}, we developed Role-Based Knowledge Generation as the initial grounding component of our prompting framework. Before producing a final response from LLMs, asking LLMs to generate potentially useful information about a given task improves the final response \cite{liu_generated_2022}. For example, as shown in an open online course "Learn Prompting" \footnote{https://learnprompting.org/}, when prompted with "\textit{Which country is larger, Congo or South Africa}?", GPT-3  answers incorrectly. However, when the model is prompted to "\textit{Generate some knowledge about the sizes of South Africa and Congo}", before answering the final question, the model uses the output to the intermediate prompt ("\textit{South Africa [has] an area of...}") to generate the correct answer: \textit{Congo is larger than South Africa}. We leverage this prompting intuition in Role-Based Knowledge Generation to enhance the model's understanding of its role as a task-performing agent. By doing so, the model can achieve better performance by accessing potentially relevant contextual information, as shown in prompts, P1 and P2 (Fig. \ref{rq1-workflow}). The corresponding outputs to P1 and P2 - O1 and O2, respectively -  are then integrated with a task-specific prompt (P3) in the following step. The role-based knowledge outputs (O1 and O2) allow the model to perform tasks more accurately given its enhanced understanding of its specific role for achieving the task. 

\textbf{Incremental Coaching}. Inspired by Chain of Thought (CoT) \cite{wei_chain--thought_2023}, Incremental Coaching is a technique within the Role-Based Incremental Coaching (RBIC) framework that involves breaking down a complex task into smaller, manageable sub-tasks as shown in P3-P5 in Fig. \ref{rq1-workflow}. The role-based agent is coached through a series of sub-tasks in a step-by-step manner, with each sub-task building upon the previous one. To implement Incremental Coaching effectively within RBIC, it is necessary to follow a logical sequence of sub-task prompts that allows the model to build understanding and confidence in performing the final task by generating incremental outputs (O3-O4). By breaking down the final task into a series of incremental sub-tasks, the role-based agent can gradually improve its comprehension of the final task to deliver a more accurate final response.  
\begin{algorithm}
\caption{Role-Based Incremental Coaching (RBIC)}
\label{alg:RBIC}
\begin{algorithmic}[1]
\REQUIRE User Input ($P$), Role-Based Agent (RBA)
\ENSURE Knowledge Base (KB) Final Task Output ($F$) 

\STATE \textbf{Comment: Role-Based Knowledge Generation}
\STATE \textbf{Input}: User Input $P$
\STATE \textbf{Output}: Intermediate knowledge $K$
\STATE $K \leftarrow \text{RBA.GenerateKnowledge}(P)$
\STATE \textbf{Update} Knowledge Base (KB) with $K$

\STATE \textbf{Comment: Incremental Coaching}
\STATE \textbf{Input}: Final Complex Task $T$, Sequence of Sub-Tasks: $T = \{S_1, S_2, \ldots, S_n\}$
\STATE \textbf{Output}: Incremental outputs $O_i = \{O_1, O_2, \ldots, O_n\}$ and Final Task Output $F$
\FOR{$i = 1$ \TO $n$}
    \STATE $O_i \leftarrow \text{RBA.Coach}(S_i, \text{KB})$
    \STATE \textbf{Update} KB with $O_i$
\ENDFOR
\STATE $F \leftarrow \text{RBA.FinalOutput}(KB,T)$
\RETURN $F$
\end{algorithmic}
\end{algorithm}

\textbf{Application of RBIC.} Here, we demonstrate the algorithmic conceptualization of the RBIC prompting framework in the context of generating gists. The essence of the Role-Based Incremental Coaching (RBIC) framework lies in its two core algorithmic components: Role-Based Knowledge Generation and Incremental Coaching, as shown in Algorithm 1. The RBIC algorithm requires the following inputs: 
\begin{itemize}
    \item User Input \(P\): The RBIC is initialized by the user input. For example, in our study, we operationalized user input as $P = (P1, P2, P3, P4A \text{ or } P4B, P5)$, as shown in Fig.\ref{rq1-workflow}.
    \item Role-Based Agent: Essentially, this can be any prompt-based LLM. For our study, we used GPT-4 as our Role-Based Agent.
\end{itemize}

Next, the RBIC algorithm will generate the following output: 
\begin{itemize}
    \item Knowledge Base (KB): The first phase of the RBIC algorithm, denoted as Role-Based Knowledge Generation, is symbolized by the function \( \text{RBA.GenerateKnowledge}(P) \). In this step, the Role-Based Agent (in our case, we use GPT-4, but this can be substituted with any prompt-based LLMs) is prompted with a user input \( P \) to elicit relevant background knowledge \( K \). This knowledge forms the basis for task execution and is stored in an initial Knowledge Base (KB).
    \begin{equation}
    K \leftarrow \text{RBA.GenerateKnowledge}(P)
    \end{equation}
    Here, \( K \) represents the knowledge generated, and \( P \) represents the user input posed by the user. \( \leftarrow \) signifies the assignment of generated knowledge \( K \) to the Knowledge Base (KB), thus creating a dynamic knowledge architecture that adapts over time. For instance, in our study, \( K \) comprised of O1 and O2 (as shown in the upper right of Fig. \ref{rq1-workflow}), which collectively formed our Knowledge Base (KB).
    \item Final Task Output (F): The subsequent phase, known as Incremental Coaching, is predicated on a sequence of sub-tasks \(\{S_1, S_2, \ldots, S_n\} \) and their corresponding outputs: 
    
    \( \{O_1, O_2, \ldots, O_n\} \). 
    \begin{equation}
    O_i \leftarrow \text{RBA.Coach}(S_i, \text{KB})
    \end{equation}

In this phase, each sub-task \( S_i \) leverages the updated Knowledge Base (KB) to produce an output \( O_i \). \( O_i \) is then used to update the KB, thus iteratively coaching the model through a series of sub-tasks in a step-by-step manner. Breaking down the final Complex Task \(T\) into simpler sub-tasks \( S_i \) allows the model to incrementally build up the necessary knowledge and skills to tackle the final task. Therefore, this incremental knowledge building across sub-tasks enables the model to better understand and perform the final Complex Task \(T\). In our case, our Complex Task ($T$) generates a "gist" based on the cause-effect pairs. The individual sub-tasks that contribute to this complex task are labeled as P3, P4A, P4B and P5 (Fig. \ref{rq1-workflow}). The algorithm proceeds sequentially, producing intermediate outputs O3, O4, and ultimately culminating in O5, which is the gist generated from the cause and effect pairs identified in sub-task P4A.
\end{itemize}

When applied to predicting gists in social media conversations, RBIC instructs the model to understand the concept of cause-effect relations as a task-performing agent. The model then incrementally performs sub-tasks to recognize and extract cause-effect pairs, and finally generates a concise gist that captures the essence of the identified causal relationship.

\begin{table*}[!htb]
\caption{Sample results from applying the RBIC method for extracting cause-effect relationships and generating gists from Reddit posts discussing health mandates between May 2020 and October 2021, including the post content}
\centering
{\fontsize{8.5}{9.9} \selectfont 
\begin{tabular}
{cL{0.25\textwidth}cL{0.12\textwidth}L{0.12\textwidth}m{0.30\textwidth}}
\hline
\rowcolor{customcolor2} \textbf{No.} & \textbf{Reddit Post} & \textbf{Label} & \textbf{Cause} & \textbf{Effect} & \textbf{Gist} \\ \hline
\rowcolor{customcolor3} 1&99.995\% of children survive cv infection, why are they pushing so hard to have kids take an experimental vaccine?& Yes & they are pushing so hard to have kids take an experimental vaccine & 99.995\% of children survive cv infection & Despite the high survival rate of children from cv infection, there is a push to have kids take an experimental vaccine.
\\ \hline
\rowcolor{customcolor4}2&Had my Pfizer jab last Wed and have felt like death since. & Yes & Had my Pfizer jab & have felt like death since & The cause of feeling like death is having the Pfizer jab last Wednesday. 
\\ \hline
\rowcolor{customcolor3}3&Imagine pointing and laughing at a single father of 3 who's now jobless and has to take care of 3 kids with no income all because he didn't want to wear a face diaper or take the experimental gene modification. & Yes & he didn't want to wear a face diaper or take the experimental gene modification & he's now jobless and has to take care of 3 kids with no income & The man's refusal to wear a face mask or take the experimental gene modification led to him losing his job and being unable to provide for his three children, resulting in financial hardship and increased responsibility for him. 
\\ \hline
\rowcolor{customcolor4}4&LA Fitness cancelled my membership against my will today because I refused to wear a mask. & Yes & I refused to wear a mask & LA Fitness cancelled my membership & The cause of LA Fitness cancelling the membership was the refusal of the person to wear a mask, which led to the effect of the membership being canceled against their will. 
\\ \hline  
\rowcolor{customcolor3}5&I’ve been thinking a lot about COVID data that’s been circulating and want to share some thoughts. I think it’s essential to remember that COVID data is not beyond skepticism, because what counts as a case varies. & Yes & what counts as a case varies & COVID data is not beyond skepticism & The variation in what is considered a COVID case has led to skepticism about the accuracy and reliability of COVID data. 
\\ \hline
\rowcolor{customcolor4}6&I stumbled upon some news. Governor Wolf has a false positive, won’t admit it because it would be admitting the tests are unreliable. And do you think it’s possible that politicians might hide their own false positive results to maintain confidence in the testing system? & Yes & false positive, won’t admit it & it would be admitting the tests are unreliable & Governor Wolf won’t admit a false positive because it would undermine COVID-19 test reliability, potentially affecting public health and safety measures. 
\\ \hline
\end{tabular} } 
\label{rq1-result}
\end{table*}

\subsubsection{Human Evaluation.}
To assess the effectiveness of RBIC's application in predicting gists in our data, we conducted a human evaluation of the RBIC-generated outputs. We recruited 6 human evaluators to evaluate the presence of causal coherence (O3), cause-effect pairs (O4), and gists (O5) for each Reddit post based on the following criteria:

\begin{itemize}
\item \textbf{Accuracy (classification)}: Is there a cause-effect relationship in the post (1/0; Yes/No)?
\item  \textbf{Relevance (extraction)}: How well does the cause-effect pair capture the primary causal relationship in the post (1-5; not well at all, slightly well, moderately well, very well, extremely well)? 
\item \textbf{Conciseness (generation)}: How well does the gist concisely summarize the cause-effect relationship in the post (1-5; not well at all, slightly well, moderately well, very well, extremely well)?
\end{itemize}

To mitigate error propagation, the evaluation was designed as a sequential process but with checks for accuracy. First, evaluators focused on `Accuracy', verifying the presence of a cause-effect relationship. Second, `Relevance' was examined to ensure the identified cause-effect pairs accurately reflected the post's main causal relationship. The final and third evaluation stage, `Conciseness', was only evaluated in posts that had already met the `Accuracy' and `Relevance' criteria. This approach minimized propagation of errors from earlier stages.

The accuracy criteria assesses the model's performance in identifying the presence of a causal relationship in a post. Relevance evaluates the model's ability to correctly extract the cause and effect phrases that are most salient to the core message of the post's content. Conciseness assesses the model's generative performance in concisely synthesizing a coherent gist based on the identified cause and effect phrases. In total, each of the 6 annotators evaluated 3,100 posts that were randomly selected from the entire dataset. For each criteria, each post received three evaluation scores from three annotators. The evaluators' assessment of the model's performance across the three criteria were generally high based on inter-rater agreement scores using Fleiss kappa ($k$) \cite{fleiss_statistical_1981}: accuracy ($k$ = $0.892$); relevance (mean = $4.3$, $k$ = $0.839$); conciseness (mean = $4.5$, $k$ = $0.864$).

\subsection{Result}
Table \ref{rq1-result} presents the results of RBIC's application, demonstrating the effectiveness of our prompting framework in predicting gists at-scale. We identified a total of 6,861 gists in our data. As shown, RBIC cannot only detect semantic causality (O3), but also extract verbatim phrases corresponding to the main cause-effect pairs (O4), and generate coherent gists (O5) based on the identified pairs. In the first example, RBIC detects sentences where causality is implied with nuance, as well as those that are more explicitly stated. Although most of the gists accurately capture the semantic essence of the causal relationship, some are more eloquent than others. For instance, the gists in examples 2 and 4 use sentence inversions, beginning with "the cause of", while others are more semantically fluid. We also performed a comparison using fine-tuned language models (BERT, RoBERTa and XLNet), as detailed in the appendix (Table \ref{model-performance}), which showed that RBIC outperformed the baseline models in extracting cause-effect pairs (O4) by 26.6\% in F1-score when comparing RBIC to the best-performing baseline model (RoBERTa with 0.814 F1-score).

\section{Study 2: How Gists Evolve Over Time}

Given the rapidly evolving public health discussions on social media, it is crucial to examine how they evolve over time \cite{hamilton_diachronic_2018, ding_same_2023}. This enables a better understanding of shifts in public opinion and emerging concerns across contentious debates around health practices like vaccinations, mask-wearing, and social-distancing \cite{gagneux-brunon_public_2022}. Hence in Study 2, we build upon our Study 1 findings to address: \textbf{What kind of gists characterize how and why people oppose COVID-19 public health practices? How do these gists evolve over time across key events?}  To answer these questions, we extract sentence embeddings from each of the gists identified from Study 1, and cluster the embeddings to identify distinct gist clusters that characterize the core topics at the center of how people argue against COVID-19 health practices. 
\renewcommand{\arraystretch}{1.0}


\subsection{Method}
\subsubsection{Extracting Sentence Embeddings from Gists.}
To identify the most salient topics across the causal language (gists) surrounding the social media discourse against public health practices, we use Sentence-BERT (S-BERT) to extract semantically rich representations of the gists identified in Study 1. S-BERT is a transformer-based model designed to produce contextualized sentence embeddings, which are particularly valuable in clustering texts  \cite{reimers_sentence-bert_2019,thakur-2020-AugSBERT}. After preprocessing the gists with standard text cleaning operations (lowercasing, removal of special characters,  tokenization), we implemented S-BERT using the SentenceTransformer to extract embeddings from our gists. The S-BERT model comprises 12 hidden layers, with each layer producing an output representation of $1 (\mathcal{N}) \times 768 (\mathcal{M})$ dimensions. To obtain high-quality embeddings, we extracted the output representations from each of the last three hidden layers of the model (layers 10-12), and computed their means. By doing so, we are able to capture and generate semantically rich representation of each gist as high dimensional vectors. 

\subsubsection{Clustering of Sentence Embeddings.}
After obtaining the sentence embeddings, we applied Principal Component Analysis (PCA) \cite{bro_principal_2014} to reduce the dimensionality of the embeddings prior to the clustering step. This was done to better visualize the language embeddings in a lower-dimensional space and to facilitate a more effective interpretation of the embedding results. We selected PCA as our method, given its frequent use and proven effectiveness in reducing dimensionality, especially for language embeddings \cite{Song2019A}. We used k-means \cite{na_research_2010} for clustering, as it is especially reliable for clustering semantic word representations \cite{yin_sentence-bert_2023}. The k-means algorithm iteratively assigns each embedding to a cluster with the closest centroid, and updates the centroid by calculating the mean of the embeddings assigned to the cluster \cite{yin_sentence-bert_2023}. This process continues until the centroids stabilize. To enhance the reliability and robustness of our clustering approach, we incorporated sentence embeddings of posts that did not contain any gists, following Samosir's study \cite{samosir_besklus_2022}. This step allows us to assess the quality of sentence embeddings by verifying that embeddings from sentences that do not contain gists cluster apart from embeddings derived from gists. Finally, we used the elbow method \cite{nainggolan_improved_2019} to determine the optimal number of clusters by calculating the sum of squared errors (SSE) in ascending order of cluster numbers until additional clusters resulted in diminishing returns \cite{marutho_determination_2018}. 

\subsubsection{Verifying Gist Clusters.}
The first author initially identified the primary themes of each cluster through categorization, screening, and summarization of 200 randomly selected gists from each cluster. Next, we recruited 6 annotators to manually evaluate and verify five primary gist clusters, as shown in Table \ref{rq2-cluster-result}. Annotators manually evaluated the clustering results by iteratively examining and discussing the themes across 200 randomly selected gists belonging to each cluster (1/0). See annotation agreement in Table \ref{IRR-table}. The verification process also includes two additional steps: (1) refining cluster descriptions such that they were thematically salient and representative of the core ideas and topics embodied by the gists in each cluster and (2) examining the sentences in the non-gist cluster that did not include any gists (non-gist cluster is \grayhl{C6}).

\begin{table*}[!h]
    \caption{Inter-rater reliability scores for human evaluation of topic clustering results. Fleiss' kappa coefficient was calculated to assess agreement between six annotators judging whether each gist was correctly assigned to one of the five clusters listed.}
    \centering
    \begin{tabular}{lc}
        \hline
         \rowcolor{customcolor2} \textbf{Cluster} & \textbf{IRR value} (Fleiss $k$)\\
         \hline
         {\redhl{C1.} Implications of Vaccine Policies, Efficacy, Side-Effects} & 0.926 \\ \hline
         {\pinkhl{C2.} Controversies Related to Masks-Wearing Practices}  & 0.894 \\ \hline
         {\greenhl{C3.} Impact of Lockdown}   & 0.884 \\ \hline
         {\bluehl{C4.} Societal and Economic (Macro) Impact of COVID-19} & 0.902 \\ \hline
         {\yellowhl{C5.} Conspiracy Theories, Domestic Politics, Foreign Countries} & 0.821 \\ \hline
         {\grayhl{C6.} Lack of distinct causal relationship or coherent gists} & 0.980 \\ \hline
    \end{tabular}

    \label{IRR-table}
\end{table*}

\subsection{Result}
\begin{figure*}[!ht]
\centering
\includegraphics[width=2.1\columnwidth]{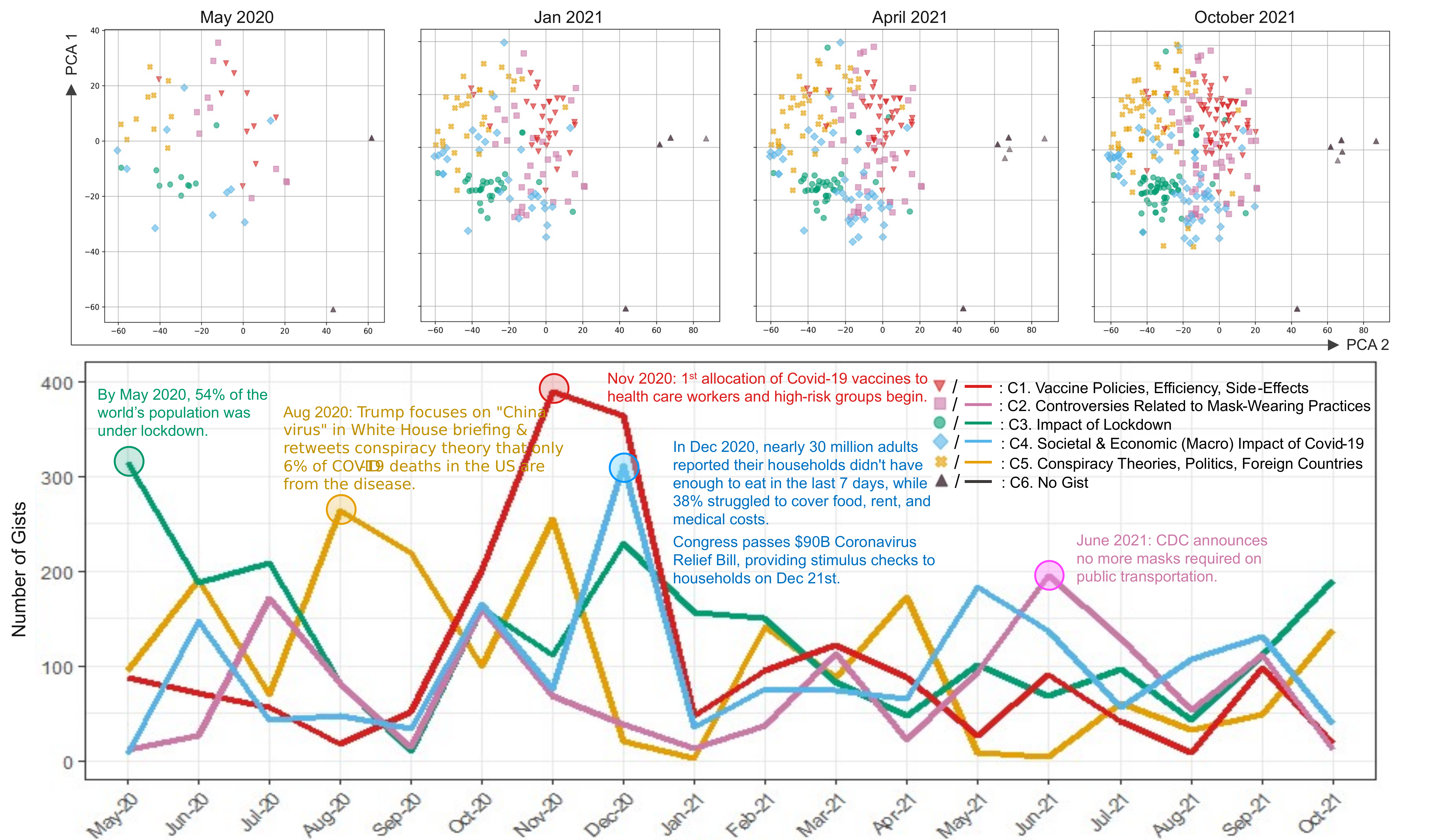}
\caption{The upper portion of the illustration displays the progression of clusters across four-month periods. The line graph illustrates the month-by-month evolution of the number of posts containing gists, representing the central themes discussed on Reddit concerning health mandates. The graph highlights specific dates when each topic was most prominently discussed and presents relevant news events related to COVID-19 and health mandates during those periods.}
\label{rq2-result2}
\end{figure*}

 A representative sample of gists from each cluster is presented in Table \ref{rq2-cluster-result}, illustrating the core topics that characterize the opposition discourse surrounding pandemic health practices. In Fig. \ref{rq2-result2},  (top panel), we visualize the evolution of our gist clusters across four time points (May 2020 - October 2021). 
 
\begin{table*}[!ht]
\caption{Representative examples are shown for each cluster from C1 to C5, highlighting the main ideas identified in the health mandate debate on Reddit. \grayhl{Cluster 6} is not included, as it lacks a distinct causal relationship or coherent gist.}
        \centering
	{\fontsize{8}{10.9}\selectfont
	\begin{tabular}{lL{0.75\textwidth}}
		\hline
		\rowcolor{customcolor2} \textbf{Cluster} & \textbf{\quad \qquad Gist}  \\ \hline
        \rowcolor{customcolor3} \multirow{5}{3.0cm}{\redhl{C1.} Implications of Vaccine Policies, Efficiency, Side-Effects}
            &   \begin{minipage}[t]{0.75\textwidth}
                    \begin{itemize}
                        \item The implementation of a vaccine mandate has resulted in people losing their jobs.
                        \item The use of experimental COVID vaccines is causing an increase in COVID deaths.
                        \item The vaccine was ineffective against new variants, which led to the death of 7,000 people who received the spike protein mRNA jab, including little kids. This suggests that the vaccine was administered for no reason, as it failed to provide protection against the new variants.
                    \end{itemize}
                \end{minipage} \\
            \hline
            \rowcolor{customcolor4} \multirow{5}{3.0cm}{\pinkhl{C2.} Controversies Related to Masks-Wearing Practices}       
           &   \begin{minipage}[t]{0.74\textwidth}
                    \begin{itemize}
                        \item If a person refuses to wear a mask at a business for medical reasons, the business may deny them services.
                        \item The lifting of mask mandates for vaccinated individuals has caused the proliferation of a deadly biohazard, which could lead to the CDC and other agencies being charged with involuntary manslaughter.
                        \item Wearing masks prevents people from seeing each other's faces, which leads to difficulties in understanding and building trust with others.
                    \end{itemize}
                \end{minipage} \\ \hline
            \rowcolor{customcolor3} \multirow{6}{3.0cm}{\greenhl{C3.} Impact of Lockdown}                                
            &   \begin{minipage}[t]{0.73\textwidth}
                    \begin{itemize}
                        \item The lockdowns have caused tourism-dependent islands in Thailand to suffer from a lack of income, leading to a situation where they have been on food aid for over a year.
                        \item The lockdowns caused a loved one to almost commit suicide, highlighting the negative impact of lockdowns on mental health.
                        \item The prolonged lockdown imposed by Cuomo for six months has resulted in the inability of the speaker to pay their bills.
                    \end{itemize}
                \end{minipage} \\ \hline
            \rowcolor{customcolor4} \multirow{5}{3.0cm}{\bluehl{C4.} Societal and Economic (Macro) Impact of COVID-19} 
            &   \begin{minipage}[t]{0.74\textwidth}
                    \begin{itemize}
                        \item The outbreak of COVID-19 has caused people to struggle with their livelihood, leading to financial difficulties and economic instability.
                        \item The COVID-19 pandemic has caused the biggest drop in US life expectancy since the second world war.
                        \item The COVID-19 shutdowns have resulted in 1 in 5 churches facing permanent closure within 18 months due to the financial strain caused by the pandemic.
                    \end{itemize}
                \end{minipage} \\ \hline
            \rowcolor{customcolor3} \multirow{8}{3.0cm}{\yellowhl{C5.} Conspiracy Theories, Domestic Politics, Foreign Countries}  
            &   \begin{minipage}[t]{0.74\textwidth}
                    \begin{itemize}
                        \item People refuse to share a table or work with certain people because they see "certain people" as sub-human because of their vaccination status.
                        \item The sentence suggests that if COVID-19 was intentionally released, it would lead to a major benefit for China and billionaires. The implication is that the cause of COVID-19's intentional release would be to bring about this benefit for these parties.
                        \item The lack of information on the epidemic from people on whether they think something is safe or not is preventing the speaker from being able to debate with their conspiracy theory friends.
                    \end{itemize}
                \end{minipage} \\ \hline
	\end{tabular}} 
\label{rq2-cluster-result}
\end{table*}

The visualization reveals interesting relations between the clusters. For instance, 
\bluehl{cluster 4}, which embodies gists discussing the impact of COVID-19 on the economy and society at-large  wraps around 
\greenhl{cluster 3} gists related to the impact of lockdown policies. This spatial proximity suggests that causal discussions on the broader consequences of the pandemic are closely intertwined with gist-based conversations on the impact of lockdown measures, shedding light on the interconnectedness of how people talk about these two topics in a causal manner.  Similarly, \redhl{clusters 1} and \pinkhl{2} are not only close in proximity, but also similar, in terms of position and shape: both clusters are diagonally positioned from top left to the bottom right, and run parallel to each other. Given that both clusters represent gists concerning specific health practices (vaccinations, mask-wearing), it is likely that these topics may share similarities in the causal manner in which people talk about the effectiveness of such health practices. \yellowhl{Cluster 5}, which represents gists related to conspiracy theories, domestic politics, and foreign countries appears to lack a clear boundary and is spatially dispersed compared to other clusters. This could be due to the fact that \yellowhl{cluster 5} encompasses multiple topics, as indicated by its description, in contrast to other clusters that are more uniformly focused on specific health practices,  government measures, or particular aspects of the pandemic's impact. The lower inter-rater reliability agreement for \yellowhl{cluster 5} (Fleiss $k$ = 0.821) further supports the notion that it is a heterogeneous cluster consisting of various topics compared to other clusters. 

The bottom half of Fig. \ref{rq2-result2} demonstrates that peak volumes of gists within each cluster align closely with key events related to the respective topics embodied by those clusters. To identify these key events, we relied on reports from major health organizations including the Centers for Disease Control and Prevention (CDC), World Health Organization (WHO), and United Nations (UN) for announcements related to public health interventions like lockdowns and vaccine rollouts \cite{wanek-libman_cdc_2021,kaplan_cdc_2020}. News reports from these organizations were widely recognized as authoritative information sources across the global community. Hence, we used such announcements and reports from these sources that highlighted key pandemic events, public health announcements, and significant milestones across the timeline of COVID-19.

In addition to these organization reports, we also analyzed articles related to COVID-19 published by major news outlets, such as AP News, Reuters, CNN, Fox News, Wall Street Journal, New York Times, and NPR. We then identified highly mediatized events by using the number of shares and article comments. This process also entailed iterative discussions among all the authors to ensure a comprehensive and balanced selection of events. Our approach aimed to minimize biases by incorporating a diverse range of sources and validating the significance of events through multiple indicators such as media coverage intensity and public engagement.

For example, \redhl{cluster 1} peak occurs in November 2020, coinciding with the country's initial phase of vaccine distributions to healthcare workers and high-risk groups \cite{kaplan_cdc_2020}. Similarly, \pinkhl{cluster 2} gists (mask-wearing) peaks in June 2021, the same month in which the federal mask mandate is lifted \cite{wanek-libman_cdc_2021}. \greenhl{Cluster 3} gists, which relate to the impact of lockdowns, peak in May 2020, by which approximately 4.2 billion or  54\% of the world's population was under lockdown \cite{May-peak-lockdown}. In December 2020, the U.S. Congress passed a bill to distribute \$90 billion in stimulus checks to households, as nearly 30 million American adults reported food and income insecurity in the same month \cite{walsh_congress_2020}. These events temporally coincide with the peak in \bluehl{cluster 4} gists, which concerns the socioeconomic consequences of the pandemic. Finally, \yellowhl{cluster 5} gists, which are related to conspiracy theories, politics, and foreign countries, reached their peak volume in August 2020, around the time when President Trump retweeted a popular online conspiracy theory \cite{Aug-peak-2} and referred to the "China virus" in his White House briefing \cite{Aug-peak-1}. Our findings imply that trends in gist volumes are linked with real-world events. 

\section{Study 3: How Social Media Gist Patterns Influence Online Engagement Behavior}
Delineating key semantic patterns (e.g., gists) that drive online behavior can help gain insight into how social media language impacts the dissemination of health information online. This, in turn, can better inform public communication strategies for time-sensitive health interventions.  Hence, in Study 3, we use Granger-causality to \textbf{examine the extent to which gist patterns influence online engagement}, such as up-voting and commenting in subreddit communities that oppose COVID-19 health practices. 

\subsection{Hypothesis Testing with Granger Causality}
Granger causality determines whether a time series $\mathcal{X}$ is meaningful in forecasting another time series $\mathcal{Y}$ \cite{granger_testing_1980}. For two aligned time series $\mathcal{X}$ and $\mathcal{Y}$, it can be said that $\mathcal{X}$ Granger-causes $\mathcal{Y}$ if past values $\mathcal{X}_{t-l} \in \mathcal{X}$ lead to better predictions of the current $\mathcal{Y}_t \in \mathcal{Y}$ than do the past values $\mathcal{Y}_{t-l} \in \mathcal{Y}$ alone, where $t$ is the time point and $l$ is the lag time or the time interval unit in which changes in $\mathcal{X}$ are observed in $\mathcal{Y}$. Lag time $l$ in Granger causality refers to the delay between a change in one time series potentially causing a change in another, indicating the time it takes for the effect to be observed. We used Granger-causality to test hypotheses 1-2, as shown below, as well as the reversed variations of H1 and H2 (H1R and H2R) where $i$ ranges from $1$ to $5$. 

\begin{center}
\begin{tcolorbox}[colback=gray!10, 
                  colframe=black, 
                  width=\columnwidth, 
                  arc=1mm, auto outer arc,
                  boxrule=0.5pt,
                  breakable, 
                 ]
\textit{\textbf{H1.}} The daily volume of gists in cluster $i$ significantly Granger-causes the upvote ratio of Reddit posts containing gists in cluster $i$.\\
\textit{\textbf{H2.}} The daily volume of gists in cluster $i$ significantly Granger-causes the number of comments associated with Reddit posts containing gists in cluster $i$.
\end{tcolorbox}
\end{center}
 
\subsection{Method and Analysis}
First, we constructed the time series data,  $\mathcal{T}_G$ for each cluster, where $\mathcal{T}_{G_i}$ represents the daily number of gists in {\textbf{cluster \textit{i}}} spanning from May 2020 to October 2021. We then created two more temporally corresponding time series data,   $\mathcal{T}_{U}$ and $\mathcal{T}_{C}$, which represent the daily upvote ratio and the daily comment count for each Reddit post containing gists from cluster $i$, respectively. We conducted a total of 20 Granger causality tests (5 clusters $\times$ 4 hypotheses - \textit{H1, H2, H1R, H2R}), using time lags ranging from 1 to 14 days.  To ensure that the value of the time series was not merely a function of time, we conducted the Augmented Dickey-Fuller (ADF) test \cite{cheung_lag_1995} using the serial difference method to achieve stationarity with ADF test values exceeding the 5\% threshold.

\begin{table*}[!ht]
\caption{Granger causality test results analyzing the relationships between the daily volume of gists in Clusters (C1-C5) and online engagement behavior - upvote ratios (UR),  number of comments (NC) -  across Reddit discussions.}
        \centering
    \begin{tabular}{llcccc}
    \hline
    \rowcolor{customcolor2} \textbf{Cluster} & \textbf{Behavior} & \textbf{Hypotheses} & \textbf{Lag (days)} & \textbf{{$F$-value}} & \textbf{$p$-value} \\
    \hline
      & upvote ratio (UR) & C1 $\not\to\mathcal{}$ UR & 4 & 4.228 & \textbf{0.022} \\
     \multirow{2}{3cm}{\redhl{C1.} Vaccination Implications} & upvote ratio (UR) & UR $\not\to\mathcal{}$ C1 & 3 & 1.739 & 0.277 \\
      & number of comments (NC) & C1 $\not\to\mathcal{}$ NC & 4 & 3.090 & \textbf{0.032} \\
      & number of comments (NC) & NC $\not\to\mathcal{}$ C1 & 7 & 1.107 & 0.445 \\
    \hline
      & upvote ratio (UR) & C2 $\not\to\mathcal{}$ UR & 6 & 5.818 & \textbf{0.012} \\
     \multirow{2}{3.4cm}{\pinkhl{C2.} Controversies and Policies Related to Masks} & number of comments (NC) & C2 $\not\to\mathcal{}$ NC & 7 & 6.007 & \textbf{0.014} \\
      & upvote ratio (UR) & UR $\not\to\mathcal{}$ C2 & 6 & 4.738 & \textbf{0.022} \\ 
      & number of comments (NC) & NC $\not\to\mathcal{}$ C2 & 9 & 0.771 & 0.463 \\
    \hline
      & upvote ratio (UR) & C3 $\not\to\mathcal{}$ UR & 2 & 0.712 & 0.545 \\
    \multirow{2}{2.9cm}{\greenhl{C3.} Impact of Lockdown} & upvote ratio (UR) & UR $\not\to\mathcal{}$ C3 & 7 & 3.410 & \textbf{0.033} \\
      & number of comments (NC) & C3 $\not\to\mathcal{}$ NC & 3 & 0.715 & 0.496 \\
      & number of comments (NC) & NC $\not\to\mathcal{}$ C3 & 8 & 6.121 & \textbf{0.016} \\
    \hline
      & upvote ratio (UR) & C4 $\not\to\mathcal{}$ UR & 3 & 6.485 & \textbf{0.011} \\
    \multirow{2}{2.9cm}{\bluehl{C4.} Societal and Economic (Macro) Impact of Covid-19} & upvote ratio (UR) & UR $\not\to\mathcal{}$ C4 & 7 & 2.011 & 0.092 \\ 
      & number of comments (NC) & C4 $\not\to\mathcal{}$ NC & 2 & 7.912 & \textbf{0.002} \\
      & number of comments (NC) & NC $\not\to\mathcal{}$ C4 & 4 & 0.815 & 0.413 \\ 
    \hline
      & upvote ratio (UR) & C5 $\not\to\mathcal{}$ UR & 5 & 1.441 & 0.512 \\
    \multirow{2}{2.9cm}{\yellowhl{C5.} Social Issues, Health, and Personal Experiences} & upvote ratio (UR) & UR $\not\to\mathcal{}$ C5 & 2 & 1.715 & 0.289 \\
     & number of comments (NC) & C5 $\not\to\mathcal{}$ NC & 4 & 2.412 & 0.089 \\
     & number of comments (NC) & NC $\not\to\mathcal{}$ C5 & 4 & 1.135 & 0.530 \\

    \hline 
    \end{tabular}
    \label{table granger}
\end{table*} 

\subsection{Results}
Table \ref{table granger} shows significant Granger causal results ($p < 0.05$). Gists across certain topics are significantly predictive of up-voting and commenting patterns, and vice-versa, in banned subreddits that oppose pandemic health practices. Specifically, the daily volume of gists significantly forecasts up-voting and commenting behavior across the topic of vaccines (\redhl{cluster 1}), mask-wearing (\pinkhl{cluster 2}),  and macro-impacts of the pandemic (\bluehl{cluster 4}) with significant lag lengths ranging from 2-7 days. These results align with prior research highlighting the linguistic power of gists in spreading online information. The reverse (H1R and H2R) is true for gists discussing the impact of lockdowns (\greenhl{cluster 3}): up-voting and commenting behavior both significantly forecast fluctuations in the volume of lockdown related gists.

\subsubsection{Bidirectional Causality:} Notably for \pinkhl{cluster 2},  which pertains to controversies and policies related to masks-wearing, we observe an interesting feedback loop between gist volumes and commenting behavior. As the volume of gists related to mask-wearing practices increases, corresponding online engagement around posts containing such gists, also increases in the form of up-votes. This behavior, in turn, further influences the volume of gists that are topically related to mask-wearing practices. In other words, there is a mutually reinforcing effect between causal language and online behavior in the context of mask-related discussions. 

\section{Study 4: How Social Media Gist Patterns Influence Nationwide Trends in Health Outcomes}
In Study 4, we address the question of \textbf{whether and how social media language patterns in the form of gists influence health decisions and outcomes in the U.S.} We follow Study 3's application of Granger causality to examine the relationship between gists patterns and important health decisions and outcomes related to COVID-19 in America. Considering the extensive attention the subreddits we analyzed received from the American public and the media \cite{NYT_Isaac_2020}, we focus on U.S. health outcomes.

\subsection{COVID-19 Data on Health Outcomes}
We used the following data from Our World in Data \footnote{https://github.com/owid/covid-19-data/tree/master/public/data}, a trusted source for COVID-19 health data for our analysis:

\begin{itemize}
    \item \textbf{Number of Vaccinations (NV):} the total number of COVID-19 vaccine doses administered on a given day.
    \item \textbf{General Hospitalization (GH):} the number of individuals hospitalized due to COVID-19 on a given day.
    \item \textbf{ICU Hospitalization (ICU):} the number of patients with COVID-19 who are in the ICU on a given day.
    \item \textbf{Total Daily COVID-19 Cases (TC):} the total number of confirmed COVID-19 cases, including probable cases.
    \item \textbf{New Daily COVID-19 Cases (NC):} the number of newly confirmed COVID-19 cases, including probable cases.
\end{itemize}

\subsection{Hypothesis Testing with Granger Causality}
Following Study 3, we Granger-test the relationship between the daily volume of gists and patterns in people's health decisions (vaccinations) and national health outcomes (General/ ICU Hospitalization, Total/ New Daily COVID-19 Cases) through H3 and its reversed variation (H3R):
\begin{center}
\begin{tcolorbox}[colback=gray!10, 
                  colframe=black, 
                  width=\columnwidth, 
                  arc=1mm, auto outer arc,
                  boxrule=0.5pt,
                  breakable, 
                 ]
\textit{\textbf{H3.}} The daily frequency of gists (Cluster $i$) significantly Granger-causes people's health decisions and/or national health outcomes, where $i$ ranges from $1$ to $5$.

\textit{\textbf{H3R.}} People's health decisions and/or national health outcomes significantly Granger-causes the daily frequency of gists (Cluster $i$), where $i$ ranges from $1$ to $5$.
\end{tcolorbox}
\end{center}

We created five time series data,  $\mathcal{T}_{NV}$, $\mathcal{T}_{GH}$, $\mathcal{T}_{ICU}$, $\mathcal{T}_{TC}$, $\mathcal{T}_{NC}$ , corresponding to the five health outcome data described above. We temporally align our data with the time frame for Studies 1-3. We performed 25 Granger causality tests (5 clusters $\times$ 5 health outcome data) with a range of lag times from 1 to 14 days. We conducted ADF tests using the serial difference method to ensure statistical robustness. 

\subsection{Results}
Table \ref{table granger_rq4} show shows significant Granger-causal results with corresponding lag lengths ($p < 0.05$). We summarize our findings below.
\begin{table*}[!ht]
\caption{Result of Granger causality test for relationships between Reddit discussion clusters (C1-C4) and health outcomes dataset. Cluster 5 is not included due to the absence of significant Granger causality findings. Notes: See Appendix A for complete statistical results (Table \ref{table granger_rq4_part1} and \ref{table granger_rq4_part2}).}
        \centering
    {\fontsize{8.5}{11}\selectfont
    \begin{tabular}{lcccccc}
    \hline
    \rowcolor{customcolor2} \textbf{Cluster} & \textbf{Health Outcomes (HO)} & \textbf{Hypotheses} & \textbf{Lag} & \textbf{$F$-val} & \textbf{$p$-val} & \textbf{Direction of Impact} \\
    \hline
    
    \multirow{2}{3cm}{\redhl{C1.} Vaccination Implications} &  Number of Vaccinations & \redhl{C1} $\not\to\mathcal{}$ NV & 4 &4.778& 0.027 \\
        & Number of Vaccinations & NV $\not\to\mathcal{}$ \redhl{C1} & 14 & 7.771 & 0.007 & \multirow{-2}{*}{bidirectional} \\

    \hline

    \multirow{2}{3.4cm}{\pinkhl{C2.} Controversies and Policies Related to Masks} & Total Daily Covid Cases & \pinkhl{C2} $\not\to\mathcal{}$ TC & 5 & 9.395 & 0.005 & \\
        & New Daily Covid Cases & \pinkhl{C2} $\not\to\mathcal{}$ NC & 5 & 11.829& 0.004 & \multirow{-2}{*}{gists impacts HO} \\
    \hline

    \multirow{2}{2.9cm}{\greenhl{C3.} Impact of Lockdown} & ICU Hospitalization & ICU $\not\to\mathcal{}$ \greenhl{C3} & 14 &4.000 & 0.039 \\
        & General Hospitalization & GH $\not\to\mathcal{}$ \greenhl{C3} & 9 &8.822& 0.006 &\multirow{-2}{*}{HO impacts gists} \\
    \hline

    \multirow{2}{3.9cm}{\bluehl{C4.} Societal and Economic (Macro) Impact of COVID-19} & Total Daily Covid Cases & TC $\not\to\mathcal{}$ \bluehl{C4} & 9 &3.663& 0.031 \\
        & New Daily Covid Cases & NC $\not\to\mathcal{}$ \bluehl{C4} & 9 &3.092& 0.032 & \multirow{-2}{*}{HO impacts gists} \\
    \hline 
    \end{tabular}}
\label{table granger_rq4}
\end{table*}

\textbf{Causal Talk Around Vaccines and National Vaccination Trends are Bidirectional}. Our results demonstrate bidirectional causality between causal discourse patterns related to vaccines and the number of vaccinations administered in the U.S. The daily volume of 
\redhl{cluster 1} gists, which consists of causal arguments related to vaccine regulations, efficacy, and side effects, is predictive of vaccination patterns across the U.S., and vice-versa. However, there is a difference in the lag lengths between H3 and H3R. It takes 4 days for gist patterns to influence vaccine adoptions (H3), while it takes two weeks for vaccination trends to shape how people talk about vaccine-related topics in a causal manner (H3R) across COVID-19 subreddits known for vaccine skepticism. In addition to a more significant Granger-causal relationship, we also observe a higher Pearson correlation for H3 ($r = 0.413$, $p = 0.005$) compared to H3R ($r = 0.105$, $p = 0.028$), indicating that national vaccination patterns have a greater impact on shaping vaccine-related causal language on social media than the other way around. There are two possible explanations: first, as more people get vaccinated, online discussions on the experiences and potential side effects of vaccines may become more prevalent - leading people to talk in a causal manner about the side effects of vaccines (e.g., "\textit{Had my Pfizer jab last Wed and have felt like death since}").  Another possible explanation is that the increasing vaccination requirements by corporations and governments as a condition for work or travel (and therefore, nationwide uptick in vaccinations) during the pandemic may have compelled vaccine-skeptics to argue more vehemently against vaccines \cite{DeDominicis2020ShoutingAE}. Previous research has shown that vaccine skeptics are susceptible to confirmation bias, as are most individuals, such that initial beliefs lead to polarization \cite{meppelink_i_2019}.  That is, vaccine skeptics are likely to seek out and discuss information about vaccines that confirms pre-existing beliefs when presented with opposing information or situated in contexts that challenge their views. Our findings align with this research, suggesting that as national vaccination uptake increases, vaccine skeptics might increasingly argue against vaccines in a causal manner (e.g., "\textit{If you take the vaccine, it's probably because you're unhealthy.}"), as commonly expressed in posts that contain \redhl{cluster 1} gists.  

\textbf{Causal Talk Around Mask-Wearing Practices Significantly Predicts Trends in COVID-19 Cases.} Our Granger-causal results show that national health outcomes, such as the total and new daily COVID-19 cases can be significantly predicted by the volume of mask-related gists (\pinkhl{cluster 2}) with a lag of 5 days.  The mask mandate was one of the most controversial health practices that impacted people of all ages and occupations during the pandemic \cite{st-amant_covid-19_2022,martin_any_2022}. Parents were polarized over school mask requirements to the extent of resorting to violence \cite{Masks-In-Schools}. Employees who asked customers to wear masks were physically assaulted \cite{bhattarai_retail_2020}. Although people initially adhered to wearing masks, more individuals started to protest mask mandates both on and offline, citing physical distress ("\textit{If having healthy lungs is important for COVID, why would we wear masks that reduce lung function?" }) or invasion of personal rights: "\textit{They will call you a `coward' or `scared' for not wanting an intrusive mask over your face (for no reason)}", as exemplified by posts containing \pinkhl{cluster 2} gists in our data. Over time, the proliferation of anti-mask views, followed by extreme resistance as demonstrated by violent altercations and wide-scale protests across the nation, may have led people to abandon mask-wearing practices \cite{grunawalt_villain_2021}, which in turn may have led to an increase in COVID-19 cases within a relatively short time-frame of 5 days, as indicated in our results.

\textbf{Rising Hospitalization Trends Prompt Causal Talk on Lockdown Impact.}  
Our findings show that nationwide trends in the number of patients hospitalized in both general and intensive care units significantly prompt more gists discussing the impact of lockdowns with a lag of 9 and 14 days, respectively (Table \ref{table granger_rq4}). Nationwide lockdowns were implemented to curb steep rises in COVID-19 cases and hospitalization rates. In fact, some posts containing \greenhl{cluster 3} gists often explicitly link lockdowns with hospitalizations: \textit{"The main reason for implementing restrictions or lockdowns was to prevent ICUs from overflowing."} Despite its necessity and intended benefit as a public health measure, studies have shown that lockdowns significantly contributed to social isolation, decrease in mental health, and rise in domestic violence across the U.S. \cite {cannon_covid-19_2021}. As the lockdown continued to amplify challenges and problems in people's lives, rising hospitalization trends across the country may have heightened people's fear and distress, leading to more intensified and causal online discourse on the lockdown's impact on everyday life. Such sentiments are clearly expressed across posts containing \greenhl{cluster 3} gists: \textit{"People are literally starting to go hungry because of lockdown restrictions"}; \textit{"The implementation of lockdowns has resulted in more harm than good"}. 

\textbf{Rising Trends in COVID-19 Cases Prompt Causal Talk on the Pandemic's Macro-Level Impact.}  
Nationwide trends in COVID-19 cases significantly Granger-causes the volume of gists discussing the pandemic's impact on society at large, with a lag of 9 days for both total and new cases. In other words, increasing trends in COVID-19 cases seem to nudge people to talk casually about the macro-level consequences of COVID-19. COVID-19 presented major economic and social setbacks that impacted all aspects of society. Some of these concerns were expressed across posts containing \bluehl{cluster 4} gists that linked the pandemic with economic crises (\textit{"The pandemic caused one of the largest economic crises, which in turn led to one of the largest poverty and hunger crises"}), decreased life expectancy (\textit{"The COVID-19 pandemic has caused the biggest drop in US life expectancy since the second world war"}), potentially oppressive public health measures (\textit{"The cause of the next deadly pandemic will lead to the implementation of authoritarian prevention measures"}), and even racism (\textit{"The fact that Covid19 affects people of color more than whites is the cause of the conclusion that Covid19 is racist"}). With COVID-19 cases rising and situations continuing to remain unpredictable, people may have become more anxious and distressed about the long-term effects on society. Consequently, this may have led individuals to discuss the pandemic's impact in a causal manner on social media, as they try to make sense of its far-reaching consequences on society \cite{Reyna2012RiskPA}.

\section{Discussion}
In summary, our findings underscore RBIC's effectiveness in efficiently predicting social media gists at scale (Study 1), thereby enriching our insight into the underlying mental constructs that shape people's health decisions and attitudes towards public health practices. In Study 2, we cluster and track the evolution of such gists, revealing key themes in online arguments against pandemic health practices. These gist volumes closely align with significant topical events, such as health announcements, policy changes, and leadership statements. In Study 3, we empirically demonstrate how gist volumes significantly drive subreddit engagement patterns (up-votes and comments). Finally, Study 4 reveals the interplay between gist patterns in anti-Covid-19 subreddits and nationwide health trends. We discuss the implications of these findings below. 

\subsection{Harnessing Large Language Models in 
Computational Social Science (CSS) Research in HCI}
Prompt-based LLMs are increasingly used in the CHI community \cite{llm-chi1,llm-chi2,llm-acl,llm-tool}, primarily contributing to the development of applications like chatbots \cite{llm-bot} and tools for co-writing \cite{llm-tool}, virtual simulations \cite{Wang2023VoyagerAO}, story-telling \cite{TaleBrush}, and visualization enhancement \cite{Singh2022ProgPromptGS}. Such studies have primarily focused on using LLMs as \textit{production} tools \cite{llm-tool} rather than tools for analysis. More recently, computational social scientists in HCI have used prompt-based LLMs for text analyses \cite{ziems_can_2023, Gilardi2023ChatGPTOC, Trnberg2023ChatGPT4OE}. However, there remain several challenges for using LLMs in nuanced examination of social media discourse. 
First, traditional NLP models and commonly used LLMs in CSS research often lack reasoning capabilities \cite{AIChains}. For instance, LLMs like BERT-based models, which are extensively used in HCI research that analyze large volumes of social media data \cite{bert-chi, Malik2023HowTD}, are typically fine-tuned for specific discrete downstream tasks (e.g., classification). While these pretrained language models have shown promise in performing discrete analyses, some emerging HCI research \cite{AIChains, Wu2022PromptChainerCL} demonstrate the additional value of prompting LLMs to perform multi-step reasoning for a more comprehensive analysis. Building on these prior insights, RBIC aims to enable a more nuanced analysis of social media discourse by leveraging the multi-reasoning capabilities of large language models. To this end, RBIC operates by performing multiple, step-by-step interrelated sub-tasks (\textit{question-answering}, \textit{classification}, \textit{extraction}, \textit{generation}) prior to generating its final output. This incremental coaching mechanism enhances the model's overall understanding and performance of the final task, allowing us to analyze social media discourse with a more comprehensive and nuanced approach.

Second, LLM development paradigms often incentivize researchers to optimize model performance using established evaluation datasets \cite{liu_generated_2022, llm-acl}. While valuable for comparing an LLM's performance with other models, this approach may not result in high performance when applied to new, unseen, in-the-wild datasets \cite{ziems_can_2023, Chung2022ScalingIL} or with tasks that are slightly different from those that the model was evaluated on \cite{ziems_can_2023}. As a result, this may limit the potential application of such LLMs for analyzing intricate, heterogeneous in-the-wild data, such as unstructured social media conversations. The role-based cognition component of RBIC addresses this limitation by allowing researchers to define and customize the role of any prompt-based LLM to perform a complex and nuanced language task. By introducing and applying RBIC in the analysis of social media conversations, we demonstrate the versatility and effectiveness of prompt-based LLMs in identifying and synthesizing nuanced linguistic patterns, thus broadening the potential application of prompt-based LLMs for theory-driven textual analysis in CSS research in the HCI domain.

\subsection{Leveraging Causal Language Patterns in Online Content Moderation Practices}

Our results show that the volume of gists across certain topics are significantly predictive of up-voting and commenting patterns, and vice-versa, in banned subreddits that oppose pandemic health practices. For example, daily gist volumes significantly predict up-voting and commenting behavior across topics related to vaccines, masks, and the pandemic's impacts, highlighting the linguistic power of gists in spreading online information as demonstrated in prior literature \cite{Valerie-pnas,reyna2023numeracy}. Similarly, our findings show that increasing trends in vaccine adoptions in the U.S. are strongly predictive of the growing volumes of vaccine-related gists in subreddits whose members are generally skeptical of vaccines. While a nationwide rise in vaccine uptake is certainly beneficial, such conditions may present challenging contexts that may reinforce vaccine skeptics to become further entrenched in their views. Vaccine opponents exposed to situations that contradict their perceptions are especially vulnerable to confirmation biases \cite{Azarpanah2021VaccineHE}, which may lead to an increased tendency to express their anti-vaccine sentiments in online communities in a causal manner, as implied by our findings.

These insights underscore the critical role of understanding and monitoring causal language patterns in public health discourse, particularly within online spaces. Current content moderation practices that rely on language models traditionally focus on flagging hate speech or monitoring specific keywords \cite{Rho2020FosteringCD}. However, our research suggests that monitoring causal language patterns can be a valuable addition to these content moderation practices, especially in controversial online communities where people exchange and learn health information. By leveraging nuanced insights from gists across various health topics, content moderation can become more effective in identifying and managing discussions that may contribute to the spread of online health misinformation or resistance to public health guidelines. 

\subsubsection{Design Implications for Moderation Dashboard:} Prior studies have shown that the design of a social media platform plays an important role in promoting transparency in content moderation \cite{Moderation-Practices}. Moderators often fail to articulate what aspect of the content prompted moderation or why such moderation was necessary \cite{Moderation-Practices}. The approach taken in our study can be built on to effectively inform users about the consequences of their posting behavior, and which aspects of their posts can potentially lead to negative outcomes. The results can also inform design strategies that  platforms can undertake to assist moderators in communicating such information to users.

Understanding and identifying causality can be difficult for humans as causality may be expressed implicitly and across sentences or intersententially \cite{Tagging}. Currently, there is no automated mechanism for moderators to systematically identify and understand the impact of causal language across online discussions. A design feature in the moderation dashboard, such as the one shown in Fig. \ref{design} (Appendix D), serves as an illustrative example of how RBIC may address this gap.  For example, when a moderator clicks on a button called ‘Enable Gist Detection (RBIC)’, an RBIC-powered extension can automatically scan posts, highlight the cause-and-effect pairs, and identify the overarching gists within the posts. This functionality may also allow moderators to see a list of top gists across community discussions in descending order of gist volumes, and an option to organize these gists based on engagement metrics, including the upvote ratio and comment volume. Additionally, the system may be designed such that the moderator may be able to drill down into posts that pertain to each of these top gists, in which the system can highlight the relevant text spans that pertain to the  cause and effect in each post.


\subsubsection{Improving Moderation and Community Guidelines} Identifying posts that do not contain moderator-specified keywords (e.g., profanity) or those that exclude explicit causal language, can still violate community norms or include misleading information in subtle ways \cite{Park2021DetectingCS}. Traditional keyword-based filters fall short in identifying such content \cite{Jhaver2019HumanMachineCF}. This can lead to difficulties in setting specific rules for moderation practices, explaining moderation-decisions, or adapting community guidelines during critical times, such as a global pandemic. With RBIC-powered gist detection, moderators can scale the searching of such posts to identify those that reflect common and theoretically predicted disconnects between the public and public health experts. This mechanism can potentially enable moderators to use concrete examples to better explain moderation decisions, as well as improve community guidelines to explain how posts that contain implicit causal narratives may impact people's knowledge and decisions around safe health practices, as shown in our work.



\subsection{Broader Implications for Understanding Engagement Patterns Across Online Communities and Offline Health Outcomes During Public Health Crises}

Our work shows that capturing psychologically important language patterns across social media, in the form of \textit{gists}, can be useful in predicting human behavior and, consequently, health outcomes. In Study 2, we demonstrate that fluctuations in the volume of gists can significantly predict online engagement patterns, specifically in terms of up-vote ratio patterns (H1) and the volume of comments (H2). This has important implications for researchers studying user behaviors in online communities \cite{survey_health_behaviors,User-Behavior}. Researchers have shown that the virality of online content is often influenced by a positivity bias in engagement metrics \cite{Kim2020TheMO}, such as up-votes and comments: posts receiving higher engagement are more visible and thus have a greater likelihood of going viral \cite{Aldous2019ViewLC, 10.1145/3173574.3174217}. This tendency can exacerbate the spread of misinformation, especially during public health crises \cite{Ahmed2022SocialMN,Silva2020PredictingMA}. Posts challenging pandemic health practices are often laden with misleading information \cite{social2023}, and online posts embedded with gists are more likely to attract more user engagement compared to those without gists \cite{broniatowski_causal_2018}. H1 and H2 results demonstrate that such user engagement patterns are predictable through gist volumes, thus highlighting the potential of using RBIC for gist analysis to track and understand the dynamics of how health-related content, especially during pandemics, resonates with and influences online user engagement. This insight is crucial for developing strategies to combat misinformation and guide public health communication effectively.

Furthermore, HCI research in crisis informatics has contributed to advancing public health monitoring systems by developing tools that track public health outcomes, online engagement patterns, or health-related topics on social media \cite{Visualizing-social,public-health}. Some of these tools that monitor online conversations extract various linguistic aspects from social media discourse, such as sentiment \cite{Visualizing-social} and topical keywords  \cite{trajkova_exploring_2020}. While these advancements have been valuable in providing descriptive insights, most do not go the full distance in linking such linguistic patterns to real-world health decisions and outcomes \cite{social2023}. Our work addresses this gap by demonstrating how RBIC can be leveraged to better connect online conversation patterns to offline health outcome trends. Study 4 results show that online causal talk related to controversial health practices, such as face-masks, are significantly predictive of total and new daily COVID-19 cases across the U.S. Likewise, our findings show that the uncertainty arising from deteriorating trends in national health outcomes may prompt people to increasingly engage online in causal discussions on the pandemic's influence on their lives and society as a whole. For example, nationwide COVID-19 cases and hospitalization patterns significantly drive up the volume of gist-based conversations concerning the pandemic's impact on society, economy, and individuals under lockdown. These findings imply that integrating gist-based language patterns into public health monitoring systems can hold promise for gaining valuable insights into the cognition that underlies skepticism and resistance to public health practices and, by extension, their impact on real-world health outcomes. Integrating RBIC-powered gist detection and real-time analysis of national health indicators into tools can potentially enhance public health agencies' ability to understand and respond to critical health challenges in relation to people's online behavior.


\section{Conclusion \& limitations}
This research synthesizes LLM techniques with theoretical perspectives from cognitive and social psychology to advance the knowledge of health decisions and outcomes in the context of the most recent pandemic. Our work is the first to systematically identify and characterize how causal language patterns surrounding anti-pandemic health practices on social media are significantly predictive of national health outcomes. These findings carry crucial implications for public health communication and policy interventions. By recognizing the influential role of causal language patterns across social media in shaping national health outcomes, public health efforts and online moderation practices can be tailored to address and mitigate the impact of social media conversations that adversely affect public health consequences.  

Our study has a limitation in our data source: it concentrates on Reddit posts and omits comments. This exclusion is primarily due to certain months of comment data being either restricted or deleted in compliance with Reddit's policies by Archive administrators. While this focus allows for an in-depth analysis of original posts, it may not capture the full discourse, including diverse viewpoints and nuanced discussions that often take place in the comments section. Consequently, our findings may offer a limited perspective on the topic under study. Future work might consider alternate ways to capture community discourse, such as through interviews or surveys, to complement the data from Reddit posts. Furthermore, as datasets from the future expand, integrating machine learning models that are capable of detecting subtle changes in discourse over time and adjust to extensive datasets may offer a dynamic view of how gists evolve. This method has the potential to uncover patterns and trends that may not be immediately obvious when using a traditional unsupervised clustering approach.

In summary, we built an LLM-based model to identify psychologically influential mental representations--gists--from social media posts, demonstrated the links between these gists and public health events, and verified associations with user engagement and national health trends, with implications for HCI design and the promotion of public health.

\bibliographystyle{ACM-Reference-Format}
\bibliography{chi}

\appendix
\clearpage

\section{Study 1 Model Performance Comparison Results}
In our study, we trained several baseline models on our human-annotated Reddit dataset to evaluate their performance in cause-effect pair extraction. Table \ref{model-performance} summarizes the performance metrics of these models.
\begin{table}[!hb]
    \caption{Model performance on cause-effect pair extraction from our manually labeled Reddit dataset.}
    \centering
    \begin{tabular}{lcccc}
        \hline
         \rowcolor{customcolor2} \textbf{Model} & \textbf{Precision} & \textbf{Recall} & \textbf{F1-score} & \textbf{Accuracy} \\
         \hline
         BERT  & 0.709 & 0.708 & 0.707 & 0.711 \\ \hline

         RoBERTa  & 0.826 & 0.829 & 0.814 & 0.846 \\ \hline
         XLNet  & 0.804 & 0.819 & 0.811 & 0.821 \\ \hline
         RBIC (\texttt{GPT-3.5})   & 0.934 & 0.949 & 0.941 & 0.940 \\ \hline
         RBIC (\texttt{GPT-4})   & \textbf{0.972} & \textbf{0.951} & \textbf{0.961} & \textbf{0.977} \\ \hline
    \end{tabular}
    \label{model-performance}
\end{table}

\section{Study 4 Statistical Result}

\begin{table*}[!hb]
\caption{Result of Granger causality test for relationships between gist clusters (C1, C2 and C3) and health outcomes dataset.}
\centering
{\fontsize{8.5}{11}\selectfont
\begin{tabular}{lcccccc}
\hline
\rowcolor{customcolor2} \textbf{Cluster} & \textbf{Health Outcomes} & \textbf{Hypotheses} & \textbf{Lag} & \textbf{$F$-val} & \textbf{$p$-val} & \textbf{Direction of Impact} \\
\hline
 \multirow{10}{3cm}{\redhl{C1.} Vaccination Implications} 
        & NV & \redhl{C1} $\not\to\mathcal{}$ NV & 4 &4.778& \textbf{0.027} & gists impacts NV \\
        & NV & NV $\not\to\mathcal{}$ \redhl{C1} & 14 & 7.771 & \textbf{0.007} & NV impacts gists \\

        & GH & \redhl{C1} $\not\to\mathcal{}$ GH & 11 &1.765& 0.271 & - \\
        
        & GH & GH $\not\to\mathcal{}$ \redhl{C1} & 5 & 1.015 & 0.410 & -  \\

        & ICU & \redhl{C1} $\not\to\mathcal{}$ ICU & 6 & 1.118 & 0.402 & -\\
        
        & ICU & ICU $\not\to\mathcal{}$ \redhl{C1} & 3 & 1.818 & 0.242 & - \\

        & TC & \redhl{C1} $\not\to\mathcal{}$ TC & 11 &2.495 & 0.089 & - \\
        
        & TC & TC $\not\to\mathcal{}$ \redhl{C1} & 11 & 2.960 & 0.072 & - \\

        & NC & \redhl{C1} $\not\to\mathcal{}$ NC & 4 &2.986& 0.068 & - \\
        
        & NC & NC $\not\to\mathcal{}$ \redhl{C1} & 9 & 0.749 & 0.488 & - \\
\hline
\multirow{10}{3.4cm}{\pinkhl{C2.} Controversies and Policies Related to Masks} 
    
        & NV & \pinkhl{C2} $\not\to\mathcal{}$ NV & 17 &0.654& 0.5124  & - \\
        & NV & NV $\not\to\mathcal{}$ \pinkhl{C2} & 7 & 0.745 & 0.4888 & - \\

        & GH & \pinkhl{C2} $\not\to\mathcal{}$ GH & 3 &1.110& 0.411 & - \\
        
        & GH & GH $\not\to\mathcal{}$ \pinkhl{C2} & 5 &2.210 & 0.091 & - \\

        & ICU & \pinkhl{C2} $\not\to\mathcal{}$ ICU & 4 &2.282& 0.090 & - \\
        
        & ICU & ICU $\not\to\mathcal{}$ \pinkhl{C2} & 14 & 1.416 & 0.343 & - \\
    
        & TC & \pinkhl{C2} $\not\to\mathcal{}$ TC & 5 & 9.395 & \textbf{0.005} & gists impacts TC \\

        & TC & TC $\not\to\mathcal{}$ \pinkhl{C2} & 8 & 1.101 & 0.414 & - \\
        
        & NC & \pinkhl{C2} $\not\to\mathcal{}$ NC & 5 & 11.829& \textbf{0.004} & gists impacts NC \\

        & NC & NC $\not\to\mathcal{}$ \pinkhl{C2} & 7 & 1.181 & 0.409 & - \\
\hline
 \multirow{10}{2.9cm}{\greenhl{C3.} Impact of Lockdown} 

        & NV & \greenhl{C3} $\not\to\mathcal{}$ NV & 4 &2.437& 0.084 & - \\
        
        & NV & NV $\not\to\mathcal{}$ \greenhl{C3} & 4 & 1.101 & 0.413 & - \\

        & GH & \greenhl{C3} $\not\to\mathcal{}$ GH & 12 &1.118& 0.410 & - \\

        & GH & GH $\not\to\mathcal{}$ \greenhl{C3} & 9 &8.822& \textbf{0.006} & GH impacts gists \\
    
        & ICU & \greenhl{C3} $\not\to\mathcal{}$ ICU & 7 &0.715& 0.494 & - \\
        
        & ICU & ICU $\not\to\mathcal{}$ \greenhl{C3} & 14 &4.000 & \textbf{0.039} & ICU impacts gists \\

        & TC & \greenhl{C3} $\not\to\mathcal{}$ TC & 4 & 1.135 & 0.409 & - \\

        & TC & TC $\not\to\mathcal{}$ \greenhl{C3} & 11 & 0.796 & 0.470 & - \\
        
        & NC & \greenhl{C3} $\not\to\mathcal{}$ NC & 5 & 2.357 & 0.088 & - \\

        & NC & NC $\not\to\mathcal{}$ \greenhl{C3} & 1 & 1.178 & 0.382 & - \\
        
    \hline
\end{tabular}
}
\label{table granger_rq4_part1}
\end{table*}

\begin{table*}[]
\caption{Result of Granger causality test for relationships between gist clusters (C4 and C5) and health outcomes dataset}
\centering
{\fontsize{8.5}{11}\selectfont
\begin{tabular}{lcccccc}
\hline
\rowcolor{customcolor2} \textbf{Cluster} & \textbf{Health Outcomes} & \textbf{Hypotheses} & \textbf{Lag} & \textbf{$F$-val} & \textbf{$p$-val} & \textbf{Direction of Impact} \\
\hline
 \multirow{10}{3.9cm}{\bluehl{C4.} Societal and Economic (Macro) Impact of COVID-19} 
    
        & NV & \bluehl{C4} $\not\to\mathcal{}$ NV & 3 &2.579& 0.072 & - \\
        
        & NV & NV $\not\to\mathcal{}$ \bluehl{C4} & 3 & 2.014 & 0.091 & - \\

        & GH & \bluehl{C4} $\not\to\mathcal{}$ GH & 9 &1.787& 0.261 & - \\

        & GH & GH $\not\to\mathcal{}$ \bluehl{C4} & 4 &2.072& 0.080 & - \\
    
        & ICU & \bluehl{C4} $\not\to\mathcal{}$ ICU & 15 &2.618& 0.066 & - \\
        
        & ICU & ICU $\not\to\mathcal{}$ \bluehl{C4} & 1 &2.071 & 0.081 & - \\

        & TC & \bluehl{C4} $\not\to\mathcal{}$ TC & 1 & 1.780 & 0.263 & - \\
        
        & TC & TC $\not\to\mathcal{}$ \bluehl{C4} & 9 &3.663& \textbf{0.031} & TC impacts gists \\
        
        & NC & \bluehl{C4} $\not\to\mathcal{}$ NC & 4 & 1.100& 0.412 & - \\
        
        & NC & NC $\not\to\mathcal{}$ \bluehl{C4} & 9 &3.092& \textbf{0.032} & NC impacts gists \\
    \hline 

    \multirow{10}{3.9cm}{\yellowhl{C5.} Social Issues, Health, and Personal Experiences} 
    
        & NV & \yellowhl{C5} $\not\to\mathcal{}$ NV & 7 &2.385& 0.085 & - \\
        
        & NV & NV $\not\to\mathcal{}$ \yellowhl{C5} & 4 & 1.101 & 0.415 & - \\

        & GH & \yellowhl{C5} $\not\to\mathcal{}$ GH & 12 &0.786& 0.471 & - \\

        & GH & GH $\not\to\mathcal{}$ \yellowhl{C5} & 14 &0.799& 0.464 & - \\
    
        & ICU & \yellowhl{C5} $\not\to\mathcal{}$ ICU & 1 &1.119& 0.399 & - \\
        
        & ICU & ICU $\not\to\mathcal{}$ \yellowhl{C5} & 9 &2.282 & 0.090 & - \\

        & TC & \yellowhl{C5} $\not\to\mathcal{}$ TC & 5 & 2.001 & 0.107 & - \\
        
        & TC & TC $\not\to\mathcal{}$ \yellowhl{C5} & 2 &1.137& 0.389 & - \\
        
        & NC & \yellowhl{C5} $\not\to\mathcal{}$ NC & 2 & 2.427& 0.079 & - \\
        
        & NC & NC $\not\to\mathcal{}$ \yellowhl{C5} & 1 &1.710& 0.212 & - \\
    \hline 
\end{tabular}}
\label{table granger_rq4_part2}
\end{table*}

\newpage
\section{Cost and Time of Running RBIC} 
\begin{table*}[]
\caption{Comparative Analysis of Cost and Time Efficiency for RBIC Framework Using GPT-4 and GPT-3.5 Models. The table provides a detailed breakdown of the estimated costs and processing time required for analyzing 79,680 posts using two different OpenAI LLMs.}
\centering{
\begin{tabular}{lllllll}
\hline
\rowcolor{customcolor2}  \textbf{Model} & \textbf{Cost per Token} (Input) & \textbf{Cost per Token} (Output)  & \textbf{Cost per Post} & \textbf{Total Cost} &  \textbf{Total Time}(hrs) \\
\hline
  GPT-4 &      \$0.03 / 1K tokens &       \$0.06 / 1K tokens &         \$0.0078 &                     \$621.50 &            199.20 \\
GPT-3.5 &    \$0.0010 / 1K tokens &     \$0.0020 / 1K tokens &      \$0.00039 &                      \$31.07 &            154.93 \\
\hline
\end{tabular}
}
\label{table-cost}
\end{table*}
Cost estimation is important for researchers considering the RBIC framework in their studies, especially with large datasets. Applying the GPT-4 pricing model of \$0.06 per 1,000 tokens (to the combined total of prompts, input posts, and outputs), the total cost for using RBIC on our dataset of 79,680 posts amounted to \$621.51, translating to roughly \$0.0078 per post. The total duration of running RBIC on GPT-4 was 199.2 hours or 8 seconds per post, as shown in Table 9. While we did not use GPT-3.5 in this study, we show the estimated cost and running time based on GPT-3.5's pricing at the time of this study. While we used GPT-4, which can be expensive for large datasets, the RBIC framework is adaptable to other open-source large language models that may be more cost-effective.

\newpage
\section{Design feature in the moderation dashboard}
\begin{figure*}[!hb]
\centering
\includegraphics[width=1.4\columnwidth]{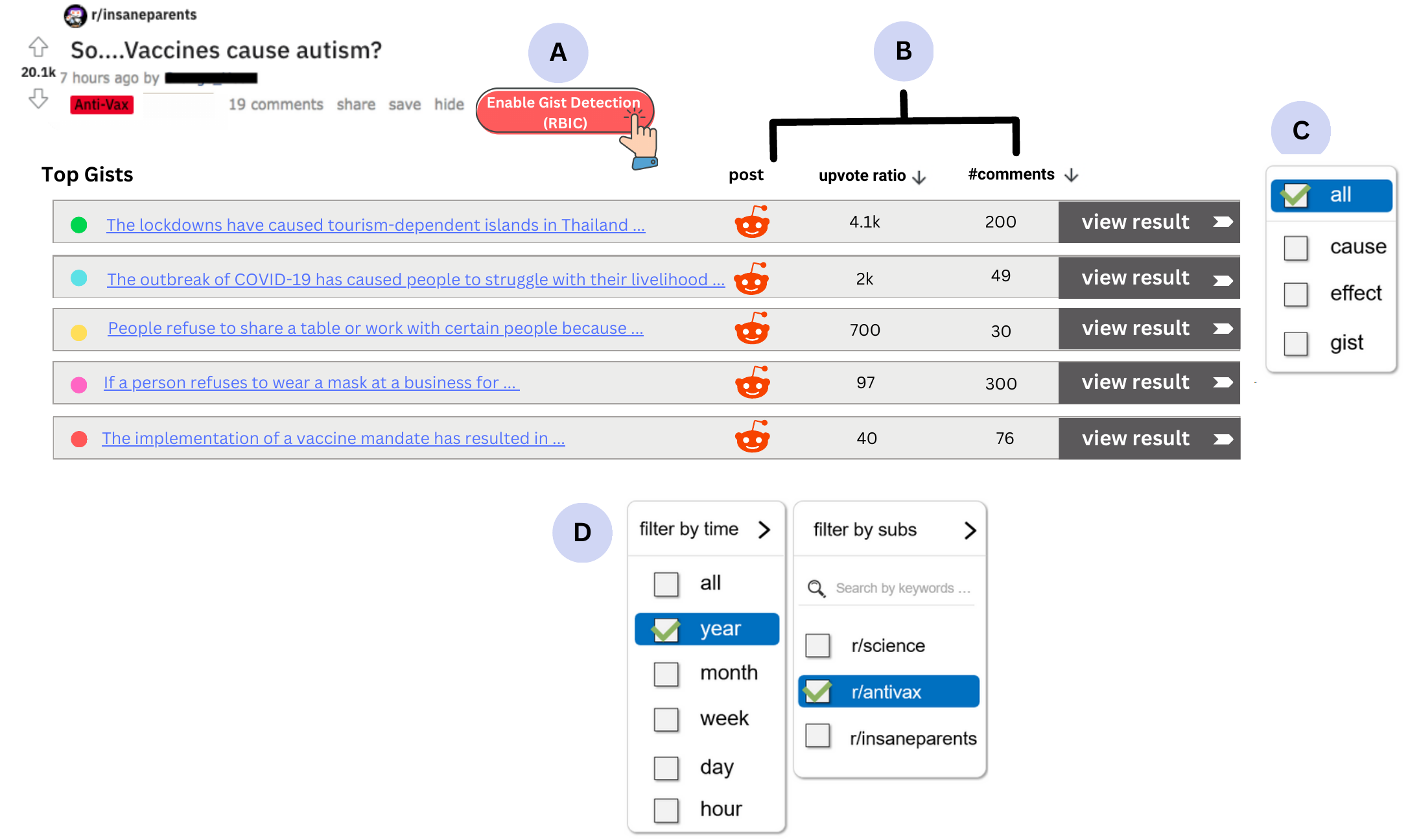}
\caption{Design Implication: Illustration of moderator’s view of a submitted post or comment.}
\label{design}
\end{figure*}


\end{document}